\documentclass[twocolumn]{aastex63}
\usepackage{epstopdf}
\usepackage{graphicx}

\shorttitle{Various Activities above Sunspot Light Bridges}
\shortauthors{Hou et al.}

\begin{document}

\title{Various Activities above Sunspot Light Bridges in \emph{IRIS} Observations: Classification and Comparison}

\correspondingauthor{Yijun Hou}
\email{yijunhou@nao.cas.cn}

\author[0000-0002-9534-1638]{Yijun Hou}
\affiliation{CAS Key Laboratory of Solar Activity, National Astronomical Observatories,
Chinese Academy of Sciences, Beijing 100101, China}
\affiliation{School of Astronomy and Space Science, University of Chinese Academy of Sciences, Beijing 100049, China}

\author[0000-0001-6655-1743]{Ting Li}
\affiliation{CAS Key Laboratory of Solar Activity, National Astronomical Observatories,
Chinese Academy of Sciences, Beijing 100101, China}
\affiliation{School of Astronomy and Space Science, University of Chinese Academy of Sciences, Beijing 100049, China}

\author[0000-0002-6565-3251]{Shuhong Yang}
\affiliation{CAS Key Laboratory of Solar Activity, National Astronomical Observatories,
Chinese Academy of Sciences, Beijing 100101, China}
\affiliation{School of Astronomy and Space Science, University of Chinese Academy of Sciences, Beijing 100049, China}

\author[0000-0002-1276-2403]{Shin Toriumi}
\affiliation{Institute of Space and Astronautical Science, Japan Aerospace Exploration Agency,
3-1-1 Yoshinodai, Chuo-ku, Sagamihara, Kanagawa 252-5210, Japan}

\author[0000-0002-9534-1638]{Yilin Guo}
\affiliation{CAS Key Laboratory of Solar Activity, National Astronomical Observatories,
Chinese Academy of Sciences, Beijing 100101, China}
\affiliation{School of Astronomy and Space Science, University of Chinese Academy of Sciences, Beijing 100049, China}

\author{Jun Zhang}
\affiliation{School of Physics and Optoelectronics engineering, Anhui University, Hefei 230601, China}

\begin{abstract}
Light bridges (LBs) are among the most striking sub-structures in sunspots, where various activities have been revealed by
recent high-resolution observations from the \emph{Interface Region Imaging Spectrograph} (\emph{IRIS}). According to the
variety of physical properties, we classified these activities into four distinct categories: transient brightening (TB),
intermittent jet (IJ), type-I light wall (LW-I), and type-II light wall (LW-II). In \emph{IRIS} 1400/1330 {\AA} observations,
TBs are characterized by abrupt emission enhancements, and IJs appear as collimated plasma ejections with a width of 1--2 Mm at
some LB sites. Most observed TBs are associated with IJs and show superpositions of some chromosphere absorption lines on
enhanced and broadened wings of C {\sc ii} and Si {\sc iv} lines, which could be driven by intermittent magnetic reconnection in
the lower atmosphere. LW-I and LW-II are wall-shaped structures with bright fronts above the whole LB. An LW-I has a continuous
oscillating front with a typical height of several Mm and an almost stationary period of 4--5 minutes. On the contrary, an LW-II
has a indented front with a height of over 10 Mm, which has no stable period and is accompanied by recurrent TBs in the entire LB.
These results support that LW-IIs are driven by frequent reconnection occurring along the whole LB due to large-scale
magnetic flux emergence or intrusion, rather than the leakage of waves producing LW-Is. Our observations reveal a highly dynamical
scenario of activities above LBs driven by different basic physical processes, including magneto-convection, magnetic reconnection,
and wave leakage.
\end{abstract}


\section{Introduction} \label{sect1}
Sunspots are the most prominent magnetic features on the Sun, which is the only star close enough to be currently observed
by humans in great detail \citep{2004ARA&A..42..517T}. As manifestations of solar magnetic field concentrations, sunspots
appear dark in solar photospheric observations because the strong magnetic fields they contain inhibit the normal
convective energy transport from the solar interior to the surface \citep{2003A&ARv..11..153S, 2011LRSP....8....4B}.
In a typical sunspot, there are an inner dark umbra and an outer less dark filamentary penumbra. It is commonly
accepted that the magnetic fields are almost vertical at the center umbra of a sunspot and become more inclined in the
surrounding penumbra \citep{2002Natur.420..390T}. In actual observations, besides the large-scale difference between the
umbra and penumbra, conditions of magnetic field and convection (thus the brightness and temperature) also change at many
sub-structures of sunspots with relatively smaller scales \citep{2003A&ARv..11..153S}. For example, in some sunspots,
bright elongated features are detected to penetrate into the dark umbra, and strong ones can even cross the umbra completely.
These features are known as light bridges (LBs) and are among the most striking sub-structures in sunspots
\citep{1969SoPh...10..384B, 1979SoPh...61..297M,1994ApJ...426..404S, 1997ApJ...484..900L}.

In high-resolution photospheric observations, sunspot LBs are usually composed of some bright ``grains" separated by a central
dark lane along the LB axis and narrow transverse intergranular dark lanes \citep{2003ApJ...589L.117B,2018ApJ...865...29Z}.
In addition, \citet{2004SoPh..221...65L} found that some LBs do not always exhibit these segmented bright grains but show
elongated filamentary structures along the length of the bridge, near whose edges small-scale barb-like features are often detected
\citep{2008ApJ...672..684R,2008SoPh..252...43L,2016A&A...596A...7S,2019ApJ...882..175Y, 2020AA...642A..44H}. It is widely accepted
that compared with the surrounding umbra, LBs have generally weaker and more inclined magnetic fields \citep{1969SoPh...10..384B,
1995A&A...302..543R,1997ApJ...484..900L}. In addition, weakly twisted and emerging magnetic fields were also found in LBs
\citep{2015A&A...584A...1L,2015ApJ...811..138T,2015ApJ...811..137T,2016A&A...594A.101Y,2020ApJ...904...84L}. In a 3D frame, forced
by the intrusion of hot and weakly magnetized plasma, the umbral magnetic fields around the LB form a magnetic canopy extending from
both sides of the LB and merging at the top \citep{2006A&A...447..343S,2006A&A...453.1079J,2014A&A...568A..60L,2016A&A...596A..59F}.
Analyzing spectral properties of a sample of 60 LBs observed with the \emph{Interface Region Imaging Spectrograph}
\citep[\emph{IRIS};][]{2014SoPh..289.2733D}, \citet{2018A&A...609A..73R} found that LBs are multi-thermal, dynamic, and coherent
structures extending up to the transition region, where diverse heating mechanisms operate in different layers.

Regarding the LB formation, \citet{2010AN....331..563S} and \citet{2012A&A...537A..19R} witnessed the formation of a thick
LB within a sunspot caused by coalescence of two pores with the same magnetic polarity \citep{1987SoPh..112...49G,2003ApJ...589L.117B}.
LBs can also form through the rapid intrusion of sunspot penumbral filaments into the umbra, accompanied by many umbral dots emerging
from the leading edges of the filaments \citep{2007PASJ...59S.577K,2021RAA....21..144L}. It is worth noting that within the special
delta-type sunspot, the collision of two rotating umbrae of opposite polarities can produce a delta-spot LB with very strong and
sheared horizontal magnetic fields \citep{1993SoPh..144...37Z,2006SoPh..239...41L,2018ApJ...852L..16O,2019ApJ...886L..21T,
2020ApJ...895..129C}.

It is obvious that the LB is a special sub-structure within a sunspot, where the magnetic field and convection conditions greatly
deviate from the ambient sunspot background. Such divergence provides a favorable environment for various dynamic activities, e.g.,
brightenings in ultraviolet (UV), G-band, and Ca {\sc ii} H channels \citep{2003ApJ...589L.117B, 2008SoPh..252...43L,
2015ApJ...811..137T, 2017A&A...608A..97F, 2020AA...642A..44H}; and jets or surges \citep{1973SoPh...28...95R, 2001ApJ...555L..65A,
2009ApJ...696L..66S, 2014A&A...567A..96L, 2016A&A...590A..57R, 2019ApJ...882..175Y}. Based on high-resolution observations from
the \emph{Hinode}, \citet{2014A&A...567A..96L} observed small-scale short-lived chromospheric jets above a sunspot LB, where the
magnetic field orientation sharply changes. It is already firmly established that this type of activities are produced by magnetic
reconnection between small-scale magnetic structures emerging in the LB and nearby umbral magnetic fields \citep{2009ApJ...696L..66S,
2016A&A...590A..57R, 2015ApJ...811..137T, 2018ApJ...854...92T, 2019ApJ...886...64Y}. Transient brightenings are usually detected
around footpoints of these jets, and intermittency is a characteristic property of these two types of activities.

In addition, recent observations from the \emph{IRIS} revealed another type of bright wall-shaped structure above sunspot LBs. Their
most prominent feature is the coherently oscillating bright front in the \emph{IRIS} 1400/1330 {\AA} channel. These structures are
called light walls (LWs) and are interpreted as the result of upward shocked p-mode waves leaked from the sub-photosphere
\citep{2015ApJ...804L..27Y, 2015MNRAS.452L..16B, 2017ApJ...838....2Z, 2017ApJ...848L...9H}. It has also been found
that external disturbances, such as flares, could change the amplitudes of LW oscillations \citep{2016ApJ...829L..29H,
2016ApJ...833L..18Y, 2017ApJ...843L..15Y}. More recently, \citet{2021ApJ...908..201L} reported small-scale bright blobs ejected
from an oscillating bright front above an LB, which together with other earlier reported activities, exposes a highly dynamical
scenario of sunspot LBs.

Although part of the various activities above sunspot LBs have been deeply investigated, due to the limitation of previous
observation condition and the lack of comparative studies, several open questions still remain: How many types of activities could
appear above sunspot LBs? What are the differences between these activities? How one type of activity is linked to another one?
What is the frequency distribution of these different activities? Insights into these questions are necessary to understand the
essential physical processes shaping the sunspot LB dynamics and call for a comprehensive and comparative investigation on the
different types of activities above sunspot LBs. In this paper, we collate these activities and analyze their different properties
mainly based on the \emph{IRIS} observations for a complete understanding of the dynamics of sunspot LBs. We report a new type of
activity above LBs, which together with other activities reported previously are classified into four distinct categories.
Furthermore, a detailed comparative study of these activities are carried out for the first time.

\section{Observations and data analysis}\label{sect2}
In this work, we mainly analyzed \emph{IRIS} and \emph{Solar Dynamics Observatory} \citep[\emph{SDO};][]{2012SoPh..275....3P}
observations over four days: 2019 May 12, 2015 January 11, 2014 October 25, and 2014 October 28. The detailed information is
summarized in Table \ref{t1}. Different types of activities above sunspot LBs are shown and compared through these observations.
In addition, \emph{IRIS} 1400/1330 {\AA} observations of sunspots in the whole year of 2014 were checked to investigate
frequency distributions of the different activities.

\begin{table*}
\caption{Summary of the Data Analyzed in This Study\label{t1}}
\centering
\begin{tabular}{c  c  c  c  c  c  c  c}   
\hline\hline
Date  & Time (UT) & Telescope  &  \emph{IRIS} Pointing (x,y) & Passbands \& Data products & Cadence & Pixel size & Figures \\
\hline
2019 May 12 & 13:54--15:21 & \emph{IRIS} & (-91{\arcsec}, 134{\arcsec}) & 1400, 2796 {\AA} & 65 s   & 0.{\arcsec}33 & \ref{fig1} \\
             &              &  &  & Raster (Very large dense 320-step) & 16.2 s  & 0.{\arcsec}35  & \ref{fig1}, \ref{fig5}, \ref{fig6} \\
             &              & \emph{SDO}/AIA  &           /               & 1700 {\AA} & 24 s & 0.{\arcsec}6 & \ref{fig1}  \\
             &              &             &                               & 171, 94 {\AA}  & 12 s & 0.{\arcsec}6 & \ref{fig1} \\
             &              & \emph{SDO}/HMI  &           /               & SHARP vector magnetic field & 720 s & 0.{\arcsec}5 & \ref{fig10} \\
             & 18:29--19:28 & \emph{IRIS} & (-39{\arcsec}, 141{\arcsec}) & 1400, 2796 {\AA} & 18 s   & 0.{\arcsec}33 & \ref{fig1} \\
             &              & \emph{SDO}/AIA  &           /               & 1700 {\AA} & 24 s & 0.{\arcsec}6 & \ref{fig1}  \\
             &              &             &                               & 171, 94 {\AA}  & 12 s & 0.{\arcsec}6 & \ref{fig1} \\
\hline
2015 Jan 11 & 03:34--04:19 & \emph{IRIS} & (-632{\arcsec}, -218{\arcsec}) & 1400, 2796, 2832 {\AA} & 38 s &  0.{\arcsec}17 & \ref{fig2}, \ref{fig8} \\
             &              &             &              & Raster (Dense synoptic)  & 9.5 s & 0.{\arcsec}35 & \ref{fig2}, \ref{fig5}, \ref{fig6} \\
             &              & \emph{SDO}/AIA  &           /               & 1700 {\AA} & 24 s & 0.{\arcsec}6 & \ref{fig2} \\
             &              &             &                               & 171, 94 {\AA}  & 12 s & 0.{\arcsec}6 & \ref{fig2} \\
             &              & \emph{SDO}/HMI  &           /               & LOS magnetogram & 45 s & 0.{\arcsec}5 & \ref{fig8} \\
             &              &                 &                           & SHARP vector magnetic field & 720 s & 0.{\arcsec}5 & \ref{fig10} \\
             & 11:41--12:36 &  \emph{IRIS} & (-611{\arcsec}, -239{\arcsec}) & 1400, 2796, 2832 {\AA} & 20 s &  0.{\arcsec}33 & \ref{fig4}, \ref{fig8} \\
             &              & \emph{SDO}/AIA  &           /               & 1700 {\AA} & 24 s & 0.{\arcsec}6 & \ref{fig4} \\
             &              &             &                               & 171, 94 {\AA}  & 12 s & 0.{\arcsec}6 & \ref{fig4} \\
             &              & \emph{SDO}/HMI  &           /               & LOS magnetogram & 45 s & 0.{\arcsec}5 & \ref{fig8} \\
             &              &                 &                           & SHARP vector magnetic field & 720 s & 0.{\arcsec}5 & \ref{fig10} \\
\hline
2014 Oct 25 & 05:17--06:30 &  \emph{IRIS} & (257{\arcsec}, -318{\arcsec}) & 1400 {\AA} & 7 s &  0.{\arcsec}17 & \ref{fig3} \\
             &              & \emph{SDO}/AIA  &           /               & 1700 {\AA} & 24 s & 0.{\arcsec}6 & \ref{fig3} \\
             &              &             &                               & 171, 94 {\AA}  & 12 s & 0.{\arcsec}6 & \ref{fig3} \\
             &              & \emph{SDO}/HMI  &           /               & LOS magnetogram & 45 s & 0.{\arcsec}5 & \ref{fig3} \\
             &              &                 &                           & SHARP vector magnetic field & 720 s & 0.{\arcsec}5 & \ref{fig10} \\
\hline
2014 Oct 28 & 08:20--09:11 &  \emph{IRIS} & (787{\arcsec}, -316{\arcsec}) & 1400 {\AA} & 39 s &  0.{\arcsec}17 & \ref{fig7} \\
             &              &              &                    & Raster (Large sparse 4-step) & 9.7 s & 1.{\arcsec}0  & \ref{fig7} \\
             & 22:16--22:58 &  \emph{IRIS} & (852{\arcsec}, -308{\arcsec}) & 1400, 2796 {\AA} & 39 s &  0.{\arcsec}17 & \ref{fig9} \\
             &              & \emph{SDO}/AIA  &           /                & 1700 {\AA} & 24 s & 0.{\arcsec}6 & \ref{fig9} \\
             &              & \emph{SDO}/HMI  &           /                & Continuum intensity & 45 s & 0.{\arcsec}5 & \ref{fig9} \\
             &              &                 &                            & SHARP vector magnetic field & 720 s & 0.{\arcsec}5 & \ref{fig10} \\
\hline
\end{tabular}
\end{table*}

The \emph{IRIS} was pointed at AR 12741 on 2019 May 12, AR 12259 on 2015 January 11, AR 12192 on 2014 October 25 and 28. Various
activities above the sunspot LBs of these ARs are clearly visible from the high-quality \emph{IRIS} observations. Here we
used Level 2 \emph{IRIS} data, including the slit-jaw images (SJIs) of 1400 {\AA}, 2796 {\AA}, and 2832 {\AA} passbands. The
emission of the 1400 {\AA} channel mainly comes from the Si {\sc iv} 1394 {\AA} and 1403 {\AA} lines formed in the transition region
($\thicksim$80,000 K) and the UV continuum emissions from the lower chromosphere. The 2796 {\AA} filter is dominated by the
Mg {\sc ii} k 2796 {\AA} line that is formed in the upper chromosphere at a temperature of $\thicksim$10,000 K. The 2832 {\AA}
passband samples the wing of Mg {\sc ii} 2830 {\AA} emissions in the upper photosphere. In addition, the \emph{IRIS} spectra in
different windows were analyzed when the \emph{IRIS} slit crossed the features of interest.

We also used observations from the Atmospheric Imaging Assembly \citep[AIA;][]{2012SoPh..275...17L} and Helioseismic and Magnetic
Imager \citep[HMI;][]{2012SoPh..275..229S} on board the \emph{SDO}. The AIA successively takes images of the solar atmosphere
in seven extreme ultraviolet (EUV) channels and three UV and visible channels. The HMI offers full-disk line-of-sight (LOS)
magnetograms, continuum intensitygrams, and photospheric vector magnetograms. Here, we analyzed the AIA 1700 {\AA}, 171 {\AA},
and 94 {\AA} images as well as the HMI LOS magnetograms, continuum intensitygrams, and vector magnetic field data product
called Space-weather HMI Active Region Patches \citep[SHARP;][]{2014SoPh..289.3549B}. The 1700 {\AA} filter mainly samples the UV
continuum emission formed around the temperature minimum region (TMR). The 171 {\AA} and 94 {\AA} passbands are dominated by the
Fe {\sc ix} 171.07 {\AA} line formed at a temperature of $\thicksim$700,000 K and the Fe {\sc xviii} 93.93 {\AA} line formed at a
temperature of $\thicksim$7,000,000 K, respectively. The HMI continuum intensitygrams and magnetograms can depict sunspots and
LBs as well as their magnetic polarity characteristics. Furthermore, the \emph{IRIS} and \emph{SDO} observations are carefully
co-aligned by matching locations of some specific features that can be simultaneously detected in different channels.

\section{Classification of Various Activities above Sunspot Light Bridges in \emph{IRIS} Observations}\label{sect3}
Based on the different physical properties and dynamic evolution of the activities above sunspot LBs observed by the \emph{IRIS},
we classified them into four distinct categories: transient brightening (TB), intermittent jet (IJ), type-I light wall (LW-I),
and type-II light wall (LW-II). The former three have been extensively studied through recent high-resolution observations, e.g.,
\emph{Hinode}, \emph{SDO}, and \emph{IRIS} observations, while the concept of LW-II is first proposed in the present work.
Note that we cannot exclude the possibilities that some rare phenomena are missed due to the gap of \emph{IRIS}
observations or more activities would be discovered in future observations. In the following four subsections, typical events
for each type of activities are shown in Figures \ref{fig1}--\ref{fig4}, where their morphological, temporal, thermal, and
spectral properties are described and investigated through \emph{IRIS} and \emph{SDO} observations. We remark that the LW-I
event shown here was reported by \citet{2015ApJ...804L..27Y} and \citet{2015MNRAS.452L..16B}, and the related Figure \ref{fig3} is
adapted from Figures 1--3 of \citet{2015ApJ...804L..27Y}.

\subsection{Transient Brightening (TB)} \label{sect31}
\begin{figure*}
\centering
\includegraphics [width=0.99\textwidth]{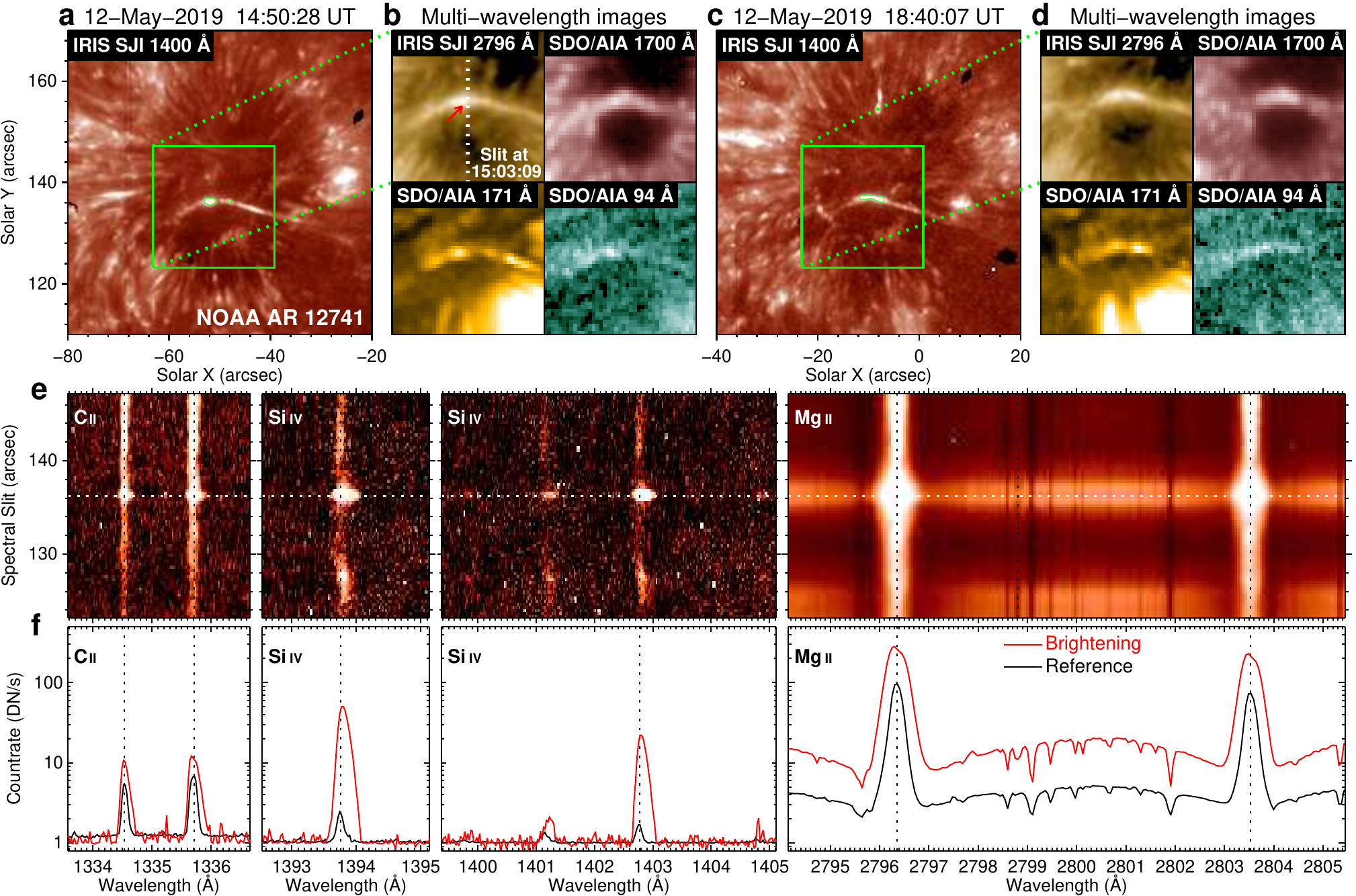}
\caption{Imaging and spectral observations of TBs in the sunspot LB of AR 12741 on 2019 May 12.
(a)--(b): Multi-wavelength images showing the brightening with a spot shape around 14:50 UT. The vertical white dotted line
in the \emph{IRIS} 2796 {\AA} image of (b) marks the position of the \emph{IRIS} scanning spectrograph slit at 15:03:09 UT when
it crossed the TB.
(c)--(d): Similar to (a)--(b), but showing the brightening with an elongated shape around 18:40 UT.
(e): \emph{IRIS} spectral detector images taken through the spectrograph slit at the position shown in (b).
(f): Spectral line profiles along the white horizontal dashed line in (e) at the brightening (red curves) indicated by the
red arrow in (b). The black line profiles represent the reference spectra obtained in a quiet plage region.
An animation of the \emph{IRIS} 1400 {\AA} and 2796 {\AA} images covering the time from 13:56 UT to 15:20 UT on 2019 May 12 is
available online. The animation's duration is 7 seconds.
}
\label{fig1}
\end{figure*}

Figure \ref{fig1} shows a typical event of TB in the LB of the main sunspot of NOAA AR 12741. This AR was located around the
solar disk center and was observed by the \emph{IRIS} several times on 2019 May 12. Here, we focus on two \emph{IRIS} data sets:
13:54 UT to 15:21 UT and 18:29 UT to 19:28 UT. The associated online animation reveals that an LB crossed the main sunspot,
where TBs appeared at different sites and sometimes repeatedly occurred at the same location. Figure \ref{fig1}(a) displays a
remarkable TB with a spot shape (see green contour) around 14:50 UT in the \emph{IRIS} 1400 {\AA} channel. The spatial scale of
this TB is approximately 1--2 Mm. Signatures of this brightening can also be clearly identified from simultaneous \emph{IRIS}
2796 {\AA}, \emph{SDO}/AIA 171 {\AA}, and 94 {\AA} observations (see panel (b)). Because these (E)UV filters mainly sample the
emission from plasma with temperatures ranging from 10,000 to 7,000,000 K, these observations indicate that the TB is
multi-thermal and that local materials could be heated to 7,000,000 K. From the animation, the brightenings could last for
several minutes. Figure \ref{fig1}(c) shows another TB with an elongated shape (see green contour) around 18:40 UT.
This TB extended along the bridge for about 3--4 Mm and lasted for several minutes in the 1400 {\AA} and 2796 {\AA} images.
In the corresponding 171 {\AA} and 94 {\AA} observations, the emission strengthening could also be identified (panel (d)).

From 13:54 UT to 15:21 UT, the \emph{IRIS} spectrograph slit scanned the region of interest from east to west in a very large
dense 320-step mode and crossed a TB in the LB around 15:03 UT. Figure \ref{fig1}(e) shows the detector images of the C {\sc ii}
1334.53/1335.71 {\AA}, Si {\sc iv} 1393.76/1402.77 {\AA}, and Mg {\sc ii} h\&k 2796.35/2803.52 {\AA} windows taken through the
slit. The white dashed line marks the TB location, along which the profiles of UV emission lines of C {\sc ii}, Si {\sc iv}, and
Mg {\sc ii} ions are measured and exemplified by the red curves in panel (f). Additionally, we averaged the spectra within a
quiet plage region around the sunspot and took the resultant profiles as reference profiles (see black curves in panel (f)).
It is obvious that compared with the reference profiles, the profiles of these spectral lines measured at the TB are enhanced
and broadened towards both the red and blue wings. Similar line profiles have recently been reported for transition region
explosive events \citep{2014ApJ...797...88H, 2015ApJ...809...82G, 2019ApJ...873...79C}, UV burst \citep{2014Sci...346C.315P,
2016ApJ...824...96T, 2018SSRv..214..120Y}, and TBs in LBs \citep{2015ApJ...811..137T, 2018ApJ...854...92T, 2020AA...642A..44H}
through \emph{IRIS} observations. In these works, such spectral line profiles were interpreted as the local heating of plasma
and bidirectional outflows derived from magnetic reconnection. Therefore, it is reasonable to suggest that the TBs in the LB and
related enhanced and broadened line profiles shown in Figure \ref{fig1} represent reconnection-related heating, which is further
supported by the observational feature of intermittency described above.

Figures \ref{fig1}(b) and (d) show that compared with other (E)UV channels, the signatures of both TBs are much weaker
in the 1700 {\AA} channel, which samples emissions mainly from the TMR in the upper photosphere. This implies that the magnetic
reconnection driving the TBs reported here might occur mainly above the TMR. Regarding the height of the reconnection, more
details of the \emph{IRIS} spectral observations will be presented and discussed in Section \ref{sect4}.

\subsection{Intermittent Jet (IJ)} \label{sect32}
\begin{figure*}
\centering
\includegraphics [width=0.99\textwidth]{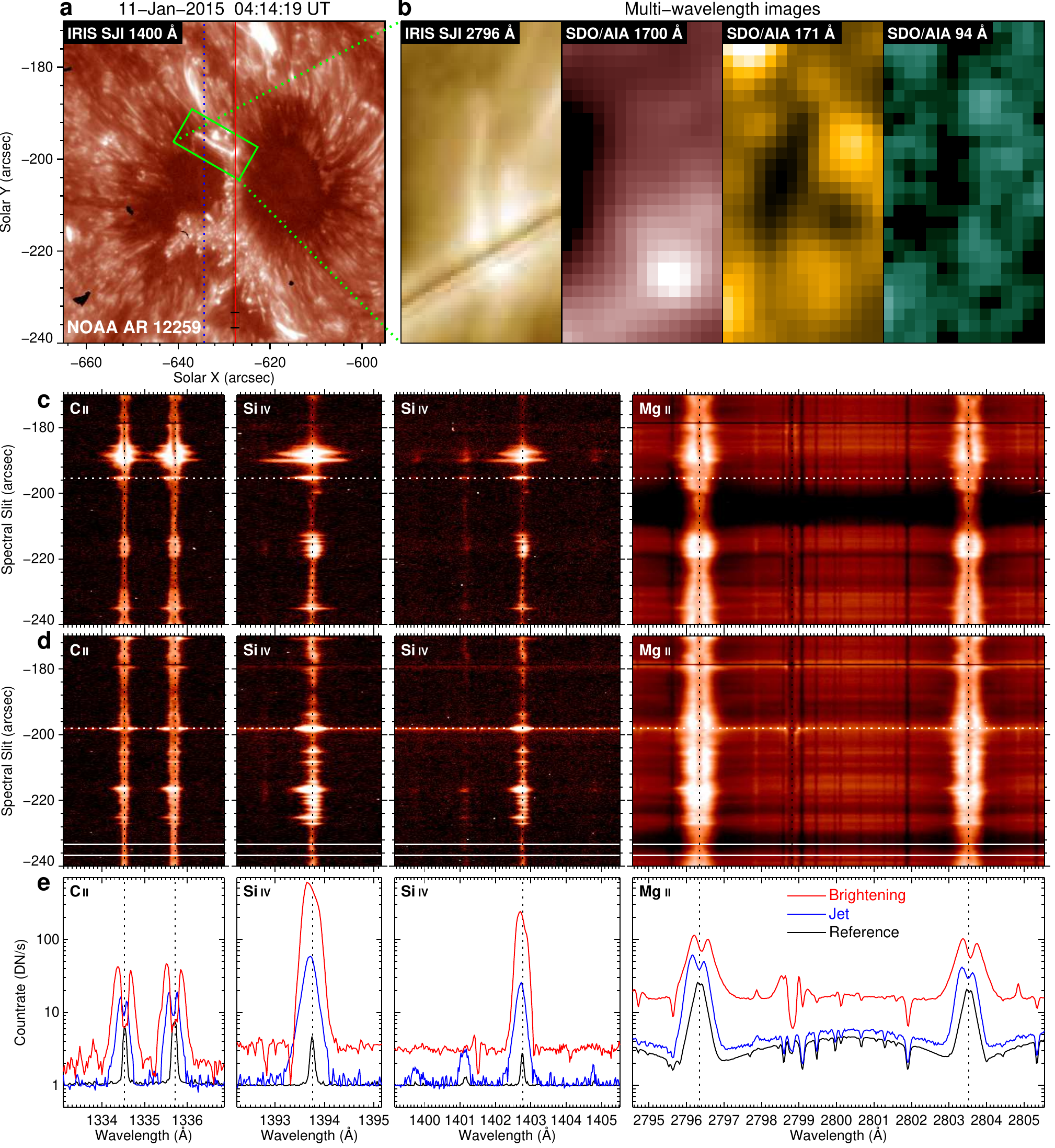}
\caption{Imaging and spectral observations of an IJ above the sunspot LB of AR 12259 on 2015 January 11.
(a)--(b): Multi-wavelength images showing the jet around 04:14 UT.
(c)--(d): \emph{IRIS} spectral detector images taken through the spectrograph slit at two positions shown by the vertical blue
and red lines in (a), respectively. The white horizontal lines mark the positions where the slit crosses the jet and
footpoint brightening.
(e): Spectral line profiles at the jet (blue curves) and its footpoint brightening (red curves). The black line profiles
represent the reference spectra averaged within the section between the two white horizontal lines in (d) (also see black
bars in (a)).
An animation of the \emph{IRIS} 1400 {\AA} and 2796 {\AA} images covering the time from 03:35 UT to 04:19 UT on 2015 January 11
is available online. The animation's duration is 7 seconds.
}
\label{fig2}
\end{figure*}

Figure \ref{fig2} exhibits an example of an IJ above the LB occurring in the main sunspot of NOAA AR 12259 on 2015 January 11.
The \emph{IRIS} observations show that jets intermittently appeared at the north end of this LB (see the associated online
animation). Figures \ref{fig2}(a) and (b) show an IJ around 04:14 UT in different filters. This IJ can be clearly identified
in emission accompanied by a footpoint brightening with a width of about 1--2 Mm in the \emph{IRIS} 1400 {\AA} and 2796 {\AA}
images. However, in the simultaneous \emph{SDO}/AIA 171 {\AA} and 94 {\AA} observations, this IJ was seen in absorption. This
difference indicates that the IJ plasma was heated to at least 80,000 K but well below 700,000 K. We note here that the
temperature of 80,000 K is only applicable under coronal ionization equilibrium and may not be used for high-density atmospheres,
where the formation temperature of Si {\sc iv} lines could be 15,000--20,000 K \citep{2016A&A...590A.124R, 2017ApJ...836...63T}.
In addition, around the tips of the IJ, enhanced emission can be dimly seen in the 1400 {\AA}, 171 {\AA}, and 94 {\AA} images,
indicating local heating that is likely due to IJ-related shock fronts or compression between the IJ and overlying atmosphere.
The jet footpoint brightening reported here is clearly visible in the \emph{SDO}/AIA 1700 {\AA} channel but dimly
visible in the 171 {\AA} and 94 {\AA} observations, which is contrary to the TBs shown in Figure \ref{fig1}. This is possibly
due to that the brightening detected here could take place in the lower atmosphere around the TMR.

From the animation, this jet reached its maximum heights of about 10 Mm within $\sim$200 s, revealing an apparent velocity of
$\sim$50 km s$^{-1}$. Considering the deceleration caused by the solar effective gravity or resistance from the upper atmosphere
as well as the projection effect, the initial true speed of the upward jets should be much higher than 50 km s$^{-1}$.
Here we estimated the Alfv\'{e}n speed ($V_{A}$) according to the formula $V_{A}=B / \sqrt{\mu_{0} \rho}$, where
$\mu_{0}$ is the magnetic field permeability of a vacuum and $\rho$ is the mass density. Assuming that the particle number
density is of the order of 10$^{15}$ cm$^{-3}$ and the magnetic field strength ($B$) is 1000 Gauss in the sunspot LB near
the TMR, the Alfv\'{e}n speed should be $\sim$70 km s$^{-1}$. The estimated Alfv\'{e}n speed around the TMR is comparable to the
lower limit (50 km s$^{-1}$) of the initial speed of the observed IJ, which suggests the possible role of magnetic reconnection
in driving the IJ reported here.

In this event, the \emph{IRIS} spectrograph slit scanned the region of interest from east to west in a dense synoptic mode and
crossed the IJ and compact brightening around the jet base. Similar to Figures \ref{fig1}(e) and (f), we show the detector images
of four spectral windows for the jet and brightening in Figures \ref{fig2}(c) and (d) and the line profiles of both features in
Figure \ref{fig2}(e). The black curves in panel (e) represent the reference profiles obtained through averaging the spectra
within a relatively quiet region marked by the black bars in panel (a). It is shown that the profiles of UV emission lines of
C {\sc ii}, Si {\sc iv}, and Mg {\sc ii} ions sampled at the jet (blue curves) all present a distinct intensity enhancement of
the blue wing, indicating excess emission contributed by moving plasma with a velocity component towards the
observer. The profiles measured at the jet footpoint brightening are notably enhanced and broadened at both wings of each
emission line, just like those of the TB shown in Figure \ref{fig1}, which represents the existence of plasma heating and
bidirectional outflows.

The above-mentioned results imply that magnetic reconnection might occur in the lower atmosphere around the TMR in this event.
After magnetic reconnection happened around the TMR, the released magnetic energy was converted into the thermal and kinetic
energy of the local plasma, which manifested as TBs and upward IJs (seen as bright jets in the warm 2796 {\AA} and 1400 {\AA}
channels but dark surges in the much hotter 171 {\AA} and 94 {\AA} channels), respectively. In addition, although not
constant and stable, the typical timescale of IJs are approximately 10--20 minutes, suggesting a convective origin. Similar
observations and interpretations have also been presented in recent works on the \emph{IRIS} observations
\citep{2015ApJ...811..137T, 2018ApJ...854...92T}.

\subsection{Type-I Light Wall (LW-I)} \label{sect33}
\begin{figure*}
\centering
\includegraphics [width=0.9\textwidth]{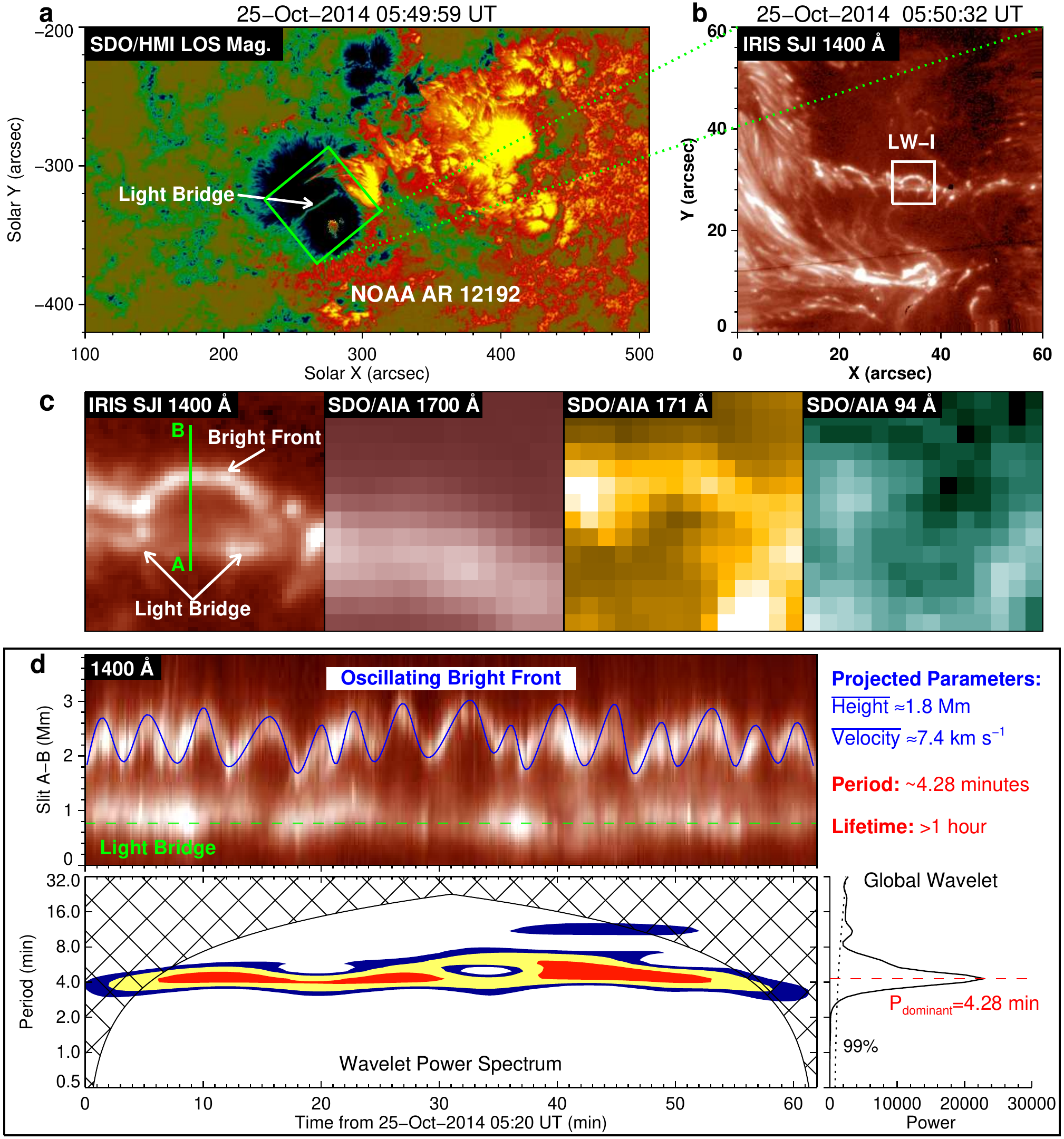}
\caption{Observations and wavelet analysis of a type-I light wall above the sunspot LB of AR 12192 on 2014 October 25.
(a): \emph{SDO}/HMI LOS magnetogram exhibiting the overview of AR 12192 and the position of the light bridge.
(b)--(c): Extended multi-wavelength images showing this type-I light wall around 05:50 UT.
(d): Time-distance plot along the slit ``A--B" in \emph{IRIS} 1400 {\AA} channel and the wavelet analysis results of the light
wall oscillations.
An animation of panel (b), covering from 05:17 UT to 06:18 UT on 2014 October 25, is available online. The animation's
duration is 10 seconds.
}
\label{fig3}
\end{figure*}

Figure \ref{fig3} shows a typical LW above the sunspot LB of NOAA AR 12192 on 2014 October 25, which was also reported by
\citet{2015ApJ...804L..27Y} and \citet{2015MNRAS.452L..16B}. Panel (a) shows an overview of AR 12192, where the
following sunspot with negative magnetic polarity was divided by a strong LB. The associated animation shows that this LW
was rooted in the whole LB and can be seen during most of the \emph{IRIS} observation period from 05:17 UT to 06:30 UT. The
field of view (FOV) outlined by the green square is rotated 141{\degr} counterclockwise and extended in panels (b) and (c)
to show this LW in different channels. The most prominent feature of this activity is a continuous bright oscillating front
detected in the \emph{IRIS} 1400 {\AA} channel. This front is also clear in the \emph{SDO}/AIA 171 {\AA} image but invisible
in the 1700 {\AA} and 94 {\AA} channels, suggesting heating of the front plasma to at least 700,000 K but well below
7,000,000 K. Between the front and wall base (i.e., LB), a bubble-like void forms. In addition, there are no obvious TBs
around the wall base, ruling out the existence of a violent release of energy.

Unlike the intermittence of the TB and IJ shown in Figures \ref{fig1} and \ref{fig2}, respectively, the LW has clear oscillating
periods of a few minutes. In Figure \ref{fig3}(d), the time-distance plot reveals long-lasting (more than one hour) and
continuous oscillations of the bright front, which is approximated by the blue curve. The distance between the LB and bright
front was calculated as the projected height of the LW, with a mean value of about 1.6 Mm; the mean rising velocity of the front
is about 7 km s$^{-1}$. Based on the time-distance plot, we applied the Morlet wavelet method to the heights of the bright front
to obtain a wavelet power spectrum, and found a period of about 4.28 minutes for the front oscillation.

The LW shows significant differences in many aspects compared with the reconnection-driven IJ. For example, the LW is rooted in
the whole LB and has a continuous bright front, which shows well-coordinated oscillations with an almost stationary period of
4.28 minutes during a relatively long time \citep[up to several days, as reported in][]{2016ApJ...829L..29H}. Moreover, no obvious
TB takes place around the base of the LW. The height of the bright front is only about 1.6 Mm, and its rising velocity is $\sim$7
km s$^{-1}$, which is much lower than the local Alfv\'{e}n speed. These observational features support the interpretation of the
LW as wave-driven activity instead of a reconnection scenario \citep{2015ApJ...804L..27Y, 2015MNRAS.452L..16B, 2017ApJ...848L...9H,
2018ApJ...854...92T}. Through analyzing \emph{IRIS} spectral data of an oscillating LW, \citet{2017ApJ...838....2Z} found
that the Mg {\sc ii} k 2796.35 {\AA} line core of the LW bright front repeatedly experiences a fast excursion to the blue
followed by a gradual shift to the red, indicating the nature of shock waves generated by the upward p-mode waves. A more detailed
comparison between the IJ and LW is made in Section \ref{sect44}.

Based on the \emph{IRIS} 1400 {\AA} observations, \citet{2016A&A...589L...7H} reported an LW with a bright front above the
umbral-penumbral boundary and found that it was rooted in an emerging magnetic field with a line-like shape. The maximum of
this LW was over 10 Mm, and recurrent brightenings were observed at the base of this LW, indicating that the reconnection
could play a key role in driving this activity. This type of LW is distinctly different from that shown in Figure \ref{fig3}
here and those reported by \citet{2015ApJ...804L..27Y,2017ApJ...843L..15Y} and \citet{2016ApJ...829L..29H,2017ApJ...848L...9H}.
In the present work, we also report such a special LW above a sunspot LB (see Figure \ref{fig4} in the following subsection).
To distinguish between these two types of LWs with different properties, LWs like that shown in Figure \ref{fig3}
are hereafter referred to as type-I LWs (LW-Is), and LWs like that shown in Figure \ref{fig4} are referred to as
type-II LWs (LW-IIs).

\subsection{Type-II Light Wall (LW-II)}\label{sect34}
\begin{figure*}
\centering
\includegraphics [width=0.88\textwidth]{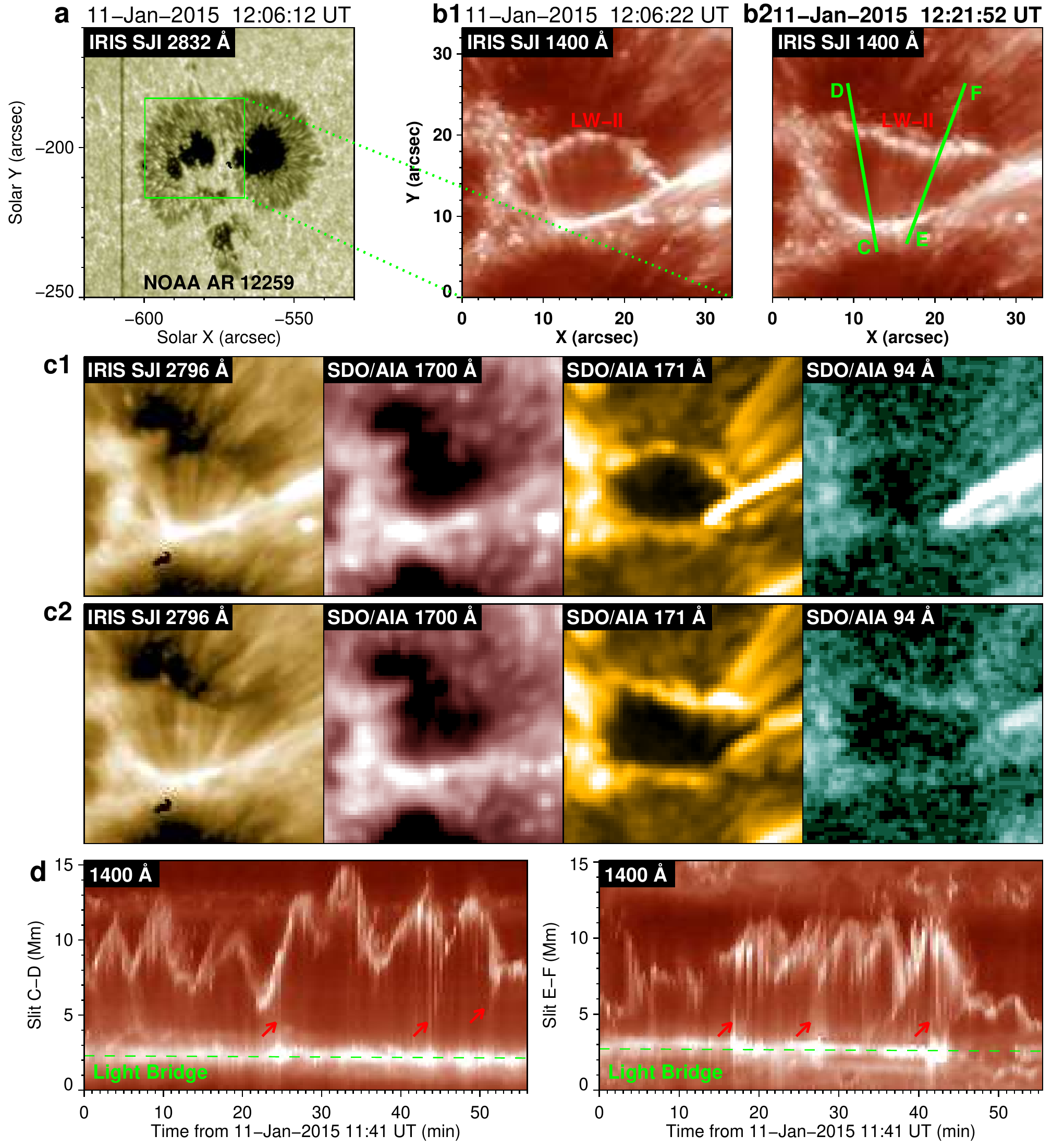}
\caption{Observations of a type-II light wall above the sunspot LB of AR 12259 on 2015 January 11.
(a)--(b2): \emph{IRIS} 2832 {\AA} image showing the sunspot LB and 1400 {\AA} images displaying the type-II light wall
above the LB at 12:06:22 UT and 12:21:52 UT, respectively.
(c1)--(c2): Multi-wavelength images showing this light wall at two time points corresponding to (b1) and (b2).
(d): Time-distance plots of 1400 {\AA} images along the slits ``C--D" and ``E--F" marked in (b2).
An animation of \emph{IRIS} 1400 {\AA} and 2796 {\AA} images, covering from 11:41 UT to 12:36 UT on 2015 January 11,
is available online. The animation's duration is 16 seconds.
}
\label{fig4}
\end{figure*}

Figure \ref{fig4} exhibits a type-II light wall above the sunspot LB of NOAA AR 12259 on 2015 January 11, which was also shown in
Figure \ref{fig2} to represent an IJ event. The associated animation reveals that this LW-II was rooted in the whole LB and had
an oscillating bright front with a height of $\sim$10 Mm in the \emph{IRIS} 1400 {\AA} channel, which is broken at some sites rather
than a continuous line. This LW-II went off in an ordered way as if one touches all piano keys in a row, or as if a sea wave
hits the harbor at an angle.
Panels (b1)--(c2) display this LW-II in different channels around 12:06 UT and 12:21 UT, respectively. The
bright front of this LW-II is also remarkable in the \emph{SDO}/AIA 171 {\AA} image but dim in the 2796 {\AA}, 1700 {\AA}, and 94
{\AA} channels, suggesting that the temperature of the front plasma is at least 700,000 K, similar to the enhanced emission at the
IJ tips. Around the base of this LW-II, recurrent brightenings were detected along the LB. In the 2796 {\AA}
and 1400 {\AA} observations, these base brightenings and the jagged bright front are connected by serried bright threads, whose
emissions are weaker than the bright front. But in the 171 {\AA} and 94 {\AA} channels, between the front and wall base, there is
a dark region similar to the IJ shown in Figure \ref{fig2}(b).

Along the two cuts ``C--D" and ``E--F" marked in panel (b2), we obtained two time--distance plots from sequences of the 1400 {\AA}
images (see panel (d)). It is shown that the bright front oscillated at two different sites but showed different patterns. Along cut
``C--D", the bright front kept oscillating during most of the period. However, along cut ``E--F", the oscillation of the front was
not continuous, and the oscillations at both sites had no stationary periods. The different performances of the LW-II above two
sites of the LB imply that the physical process dominating the LW-II can not be the leaked p-mode wave that shapes the whole LB
and drives the coherent oscillation of the LW-I. Note that plasma ejections were frequently launched from the brightenings in the
LB (see red arrows in panel (d)). The trajectories of these ejections appear to be nearly vertical in the time--distance diagrams,
indicating that these ejections reach the bright front very quickly. In successive \emph{IRIS} observations, these ejections could
correspond to the serried bright threads connecting the base brightenings and the bright front.

The results mentioned above reveal that although the LW-II is similar to the LW-I in terms of the oscillating bright front, these two
types of LWs are distinctly different in many other ways: (1) The bright front of the LW-II is broken or indented rather than a continuous
one like that of the LW-I. (2) Compared to LW-Is with typical heights of 2--5 Mm, the LW-IIs have much larger heights of $\sim$10 Mm.
(3) Around the base of an LW-I, there are usually no obvious TBs, but they recurrently appear in the whole bridge for an LW-II. (4)
Between the bright front and the LB, a bubble-like void forms in an LW-I. However, the main body of an LW-II consists of serried
bright threads caused by the fast upward ejections in the 1400 {\AA} channel, which appears as a dark region in the 171 {\AA} and
94 {\AA} channels. (5) An LW-I rises with a projected velocity of $\sim$7 km s$^{-1}$ while the LW-II has a rising velocity of
$>$50 km s$^{-1}$. (6) An LW-I performs coherent oscillation with a nearly stationary oscillating period of 4--5 minutes, while the
oscillation of an LW-II at two sites shows different patterns without constant period. In terms of these aspects, an LW-II actually
resembles an IJ. As a result, it is reasonable to speculate that an LW-II could be dominated by magnetic reconnection over a certain
spatial range, and the indented bright front might be caused by local plasma heating due to reconnection-related shock fronts or
compression between the upward ejections and overlying atmosphere. In Section \ref{sect4}, the TB, IJ, and two types of LWs are
analyzed and compared with each other.

\section{Comparison of Various Activities above Sunspot Light Bridges in \emph{IRIS} Observations}\label{sect4}
In the following subsections, we first compare these four types of activities pairwise. There are 6 comparisons in total:
TB vs. IJ, TB vs. LW-I, TB vs. LW-II, IJ vs. LW-I, IJ vs. LW-II, and LW-I vs. LW-II. Because some groups of comparisons are closely
related to each other, we discuss them together out of the order mentioned above. Meanwhile, these activities reported here are
also compared with those observed previously by other observatories, such as \emph{Hinode}. Possible connections between the
phenomena observed by different observatories at different ages are discussed. In the end of this Section, we further compare the
four types of activities from the aspect of magnetic field and frequency distribution.

\subsection{TB Versus IJ}\label{sect41}
\begin{figure*}
\centering
\includegraphics [width=0.95\textwidth]{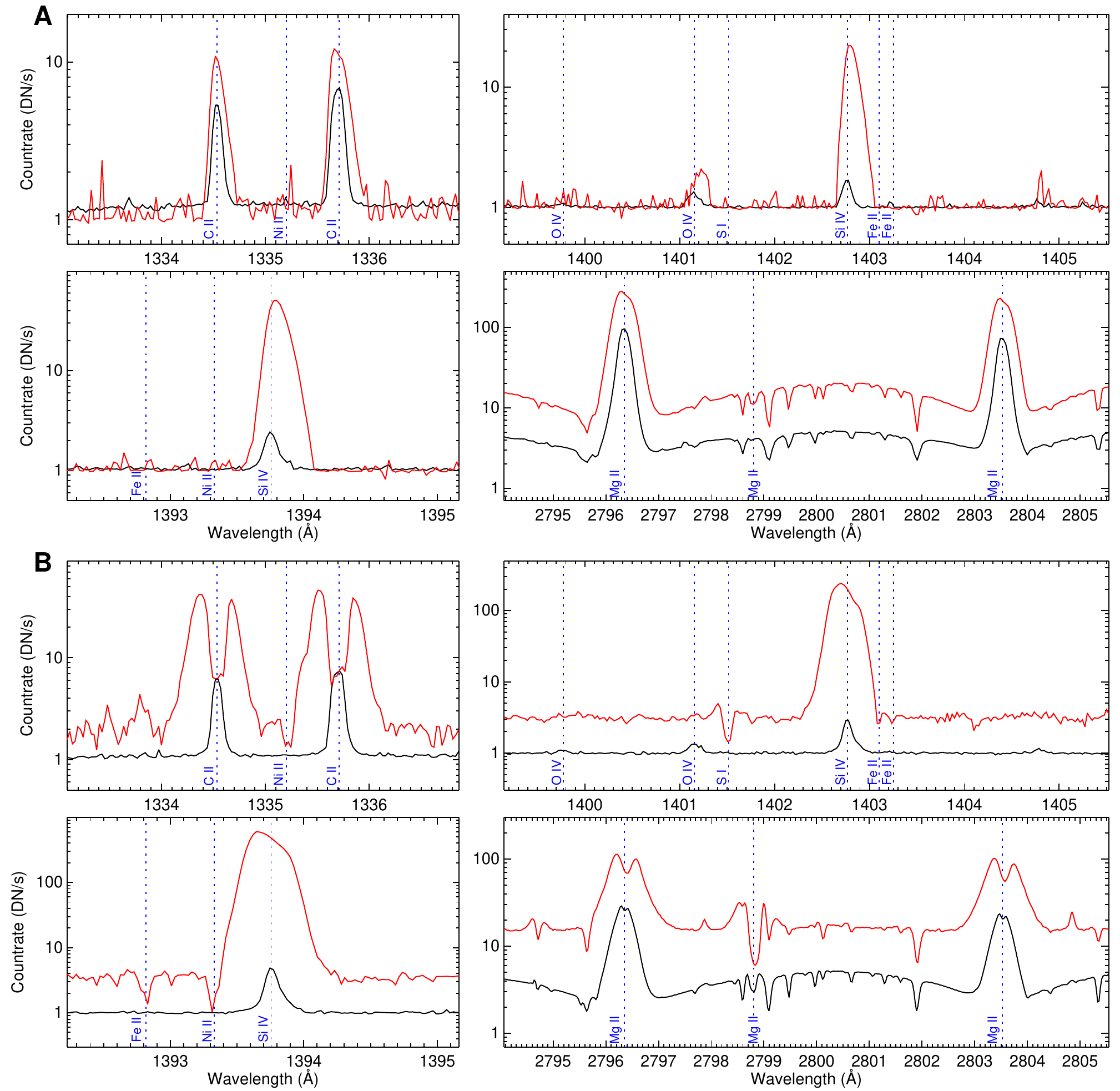}
\caption{Typical \emph{IRIS} line profiles (red curves) of two types of TBs in four spectral windows. A is for the TB without
the jet as shown in Figure 1. B is for the brightening observed at the footpoint of the jet shown in Figure 2. The black line
profiles are the reference spectra. The rest wavelengths of some lines of interest are indicated by vertical blue dashed lines.
}
\label{fig5}
\end{figure*}

TB and IJ are the most common activities observed in sunspot LBs and have been extensively investigated in many previous works.
Generally, an IJ is accompanied by a TB around its footpoint, which is interpreted as the result of magnetic reconnection
occurring in the low solar atmosphere \citep{2001ApJ...555L..65A, 2009ApJ...696L..66S, 2014A&A...567A..96L, 2016A&A...590A..57R,
2019ApJ...870...90B}. As discussed in Section \ref{sect32}, after the occurrence of reconnection, the released magnetic energy
will heat and accelerate local plasma. As a result, TBs appear in the lower atmosphere, where IJs are launched into the higher
layer. However, as shown in Figure \ref{fig1}, not all TBs are associated with IJs. Similar observations have also been presented
in \citet{2020AA...642A..44H}, where the magnetic topology and occurrence height of the reconnection (and thus the plasma density)
were considered as key factors. These two kinds of TBs (associated with IJs or not) show interesting similarities and differences,
both of which are closely related to the magnetic reconnection. To further explore the mechanisms driving these two types of TBs,
we take the two events shown in Figures \ref{fig1} and \ref{fig2} as samples and compare their multi-wavelength imaging and
\emph{IRIS} spectral observations in detail.

The TBs shown in Figure \ref{fig1} are clearly observed in the 2796 {\AA}, 1400 {\AA}, 171 {\AA}, and 94 {\AA} channels but are
weaker in 1700 {\AA} images, suggesting that these TBs are located above the TMR and heated up to coronal temperatures. For the
TB exhibited in Figure \ref{fig2}, the related signatures are obvious in the 2796 {\AA} and 1400 {\AA} channels and are particularly
strong in the 1700 {\AA} channel. However, in the hotter 171 {\AA} and 94 {\AA} images, the TB is absent. Besides, this TB is
associated with an IJ with a velocity comparable to the local Alfv\'{e}n speed. Thus, we suggest that the TB shown in Figure
\ref{fig2} is located around the TMR and is heated to only transition region temperature.

Typical line profiles of these two types of TBs are shown in Figures \ref{fig5}(A) and (B), respectively. Both TBs show enhanced and
broadened profiles of the C {\sc ii}, Si {\sc iv}, and Mg {\sc ii} lines, which are usually suggested to be related to reconnection
heating and outflows \citep{2014Sci...346C.315P, 2015ApJ...811..137T, 2018SSRv..214..120Y}. The most prominent feature in the spectra
of a TB associated with a jet is the superposition of some absorption lines on the greatly broadened wings of the C {\sc ii} and
Si {\sc iv} line profiles (see panel (B)). These lines include Ni {\sc ii} 1335.203 {\AA}, Fe {\sc ii} 1392.817 {\AA}, Ni {\sc ii}
1393.33 {\AA}, Fe {\sc ii} 1403.101 {\AA}, and Fe {\sc ii} 1403.255 {\AA}, which are typically formed in the upper chromosphere. In
addition, these absorption lines reveal a slight blueshift. It is well established that these absorption lines with slight blue
shifts are the result of upward expanding cold chromosphere material overlying a hot reconnection process (with a temperature up to
80,000 K) in the lower atmosphere \citep{2014Sci...346C.315P, 2016ApJ...824...96T}. Another remarkable feature of this TB is
that the Mg {\sc ii} 2798.809 {\AA} line changes from absorption to emission, implying the existence of lower chromosphere heating
\citep{2015ApJ...806...14P, 2020AA...642A..44H}. Moreover, the O {\sc iv} 1399.77 {\AA} and 1401.16 {\AA} lines are absent for this TB,
which implies that the TB is in a high-density environment \citep{2014Sci...346C.315P, 2016ApJ...824...96T, 2018ApJ...854...92T}.
These spectral features further reinforce our conclusion that the TB associated with the jet shown in Figure \ref{fig2} was produced
by magnetic reconnection occurring in the lower atmosphere around the TMR.

\begin{figure*}
\centering
\includegraphics [width=0.99\textwidth]{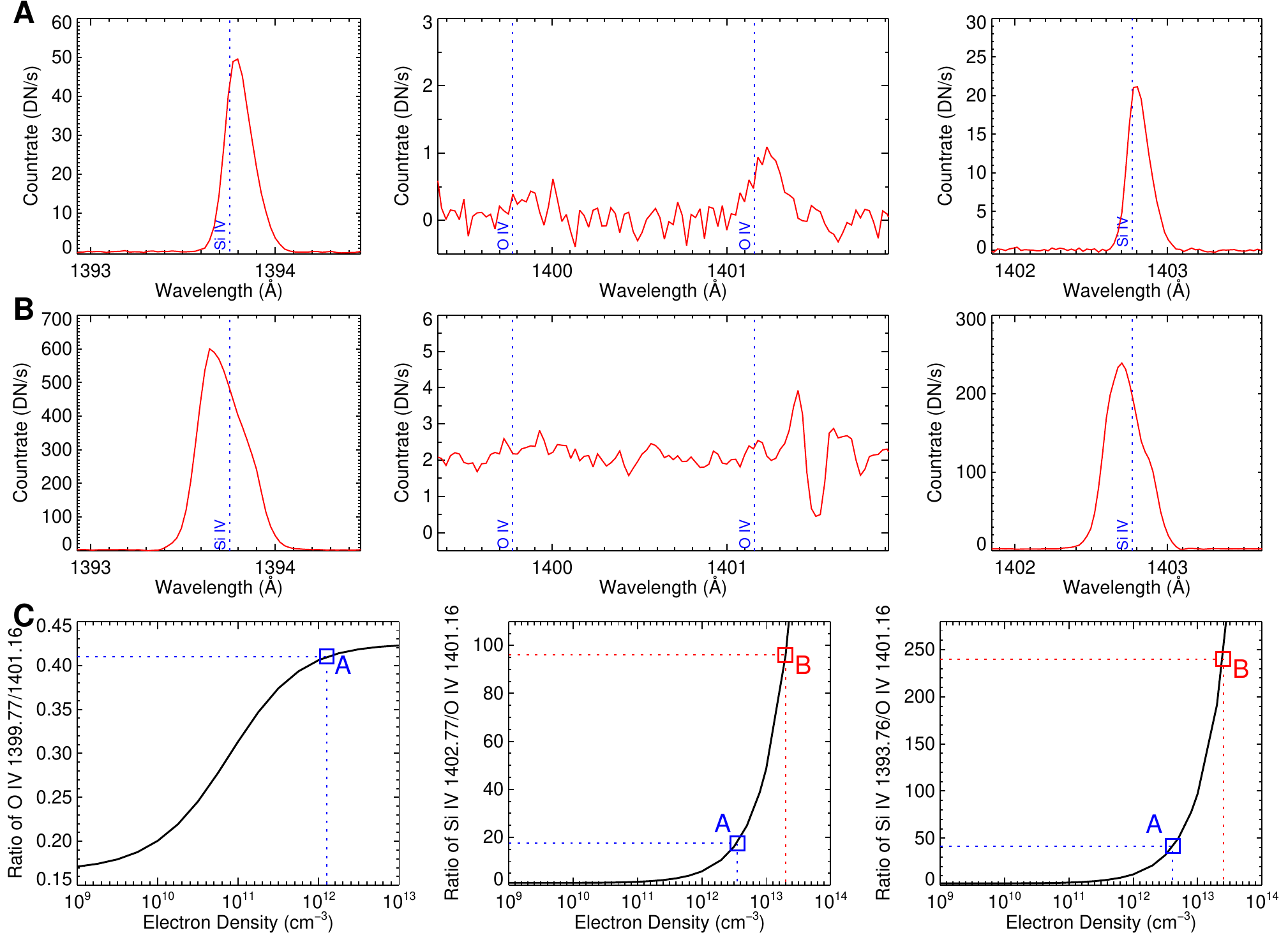}
\caption{Density diagnostics for the two types of TBs based on \emph{IRIS} spectral observations.
(A): Line profiles of Si {\sc iv} 1393.76 {\AA}, O {\sc iv} 1399.77/1401.16 {\AA}, and Si {\sc iv} 1402.77 {\AA} for the TB
without a jet shown in Figure \ref{fig1}.
(B): Similar to (A) but for the jet footpoint brightening shown in Figure \ref{fig2}.
(C): Theoretical relationships between electron density and line intensity ratios of O {\sc iv} 1399.77/1401.16 {\AA}, Si {\sc iv}
1402.77/O {\sc iv} 1401.16 {\AA}, and Si {\sc iv} 1393.76/O {\sc iv} 1401.16 {\AA}. The blue and red squares indicate the line
intensity ratios and electron densities derived from line profiles shown in A and B, respectively.
}
\label{fig6}
\end{figure*}

On the contrary, as shown in Figure \ref{fig5}(A), the spectra of the TB without a jet do not reveal the obvious signatures mentioned
above, e.g., superpositions of chromospheric absorption lines, change of the Mg {\sc ii} 2798.809 {\AA} line, and absence of the
O {\sc iv} 1399.77 {\AA} and 1401.16 {\AA} lines. Here, we should mention the possibility that the relatively weaker enhancement
and broadening of the C {\sc ii} and Si {\sc iv} line profiles could make the chromospheric absorption lines difficult to identify.
However, to a certain extent, the normal profiles of the Mg {\sc ii} 2798.809 {\AA}, O {\sc iv} 1399.77 {\AA}, and 1401.16 {\AA}
lines here could rule out the possibility that this TB without a jet is located in the lower atmosphere. Moreover, it is shown that
two Si {\sc iv} line profiles show obvious redshifts, which could be caused by the downflows from the reconnection region around
the transition region. Similar observations were found in the transition region explosive events around the footpoints of network
jets \citep{2019ApJ...873...79C, 2021ApJ...918L..20H}. Naturally, one may wonder why the TB detected here did not produce a jet.
We propose that because the TB reported here is located above the sunspot LB, the plasma density around the TB is much lower than
the environment hosting network jets. As a result, the reconnection outflows could barely be detected as jets. Another possible
explanation for the absence of jets might be their direction towards the observer along the LOS, but this possibility is ruled out
by the redshift of the Si {\sc iv} line profiles.

Furthermore, through the density diagnostics based on intensity ratios of different lines, we can obtain the electron densities
of the two types of TBs, thus inferring their heights. As shown in Figure \ref{fig6}(A), Si {\sc iv} 1393.76 {\AA}, O {\sc iv}
1399.77/1401.16 {\AA}, and Si {\sc iv} 1402.77 {\AA} can be identified from the spectra of the TB without a jet. Therefore, we
derived the electron density based on the line intensity ratios of O {\sc iv} 1399.77/1401.16 {\AA}, Si {\sc iv} 1402.77/O
{\sc iv} 1401.16 {\AA}, and Si {\sc iv} 1393.76/O {\sc iv} 1401.16 {\AA}, where the theoretical relationship between the electron
density and intensity ratio was obtained under the assumption of ionization equilibrium \citep{2014Sci...346C.315P,
2015A&A...582A..56D, 2018ApJ...858L...4X, 2019JGRA..124.9824C, 2020ApJ...897..113H}. However, for the TB with a jet (panel (B)),
the O {\sc iv} 1399.77 {\AA} and 1401.16 {\AA} lines are both weak, so their line intensity ratio is unreliable. Thus, we just
estimate its electron density based on the line intensity ratios of Si {\sc iv} 1402.77/O {\sc iv} 1401.16 {\AA} and Si {\sc iv}
1393.76/O {\sc iv} 1401.16 {\AA}. Note that due to the uncertainty of the observed line intensity and the theoretical
relationship itself, the electron density derived here should only be used for a qualitative comparison between the two types of TBs.
As shown in Figure \ref{fig6}(C), the electron density of the TB without a jet is in the range of 10$^{12}$--10$^{13}$ cm$^{-3}$,
which is significantly lower than that of the IJ-related TB  ($>10^{13}$ cm$^{-3}$). These results, together with the features
discussed above, suggest that the fundamental difference between the IJ- and non-IJ-related TBs may be the formation height: the
former are produced by reconnection in the lower atmosphere around the TMR, and the latter are produced by reconnection in the
higher transition region.

According to existing observations and studies, IJ-related TBs are dominant in sunspot LBs and are suggested to be driven by the
reconnection between small-scale emerging magnetic arcades in the LB and surrounding umbral near-vertical fields. Due to the
constraint of the magnetic canopy above the LB, the emerging fields in the bridge can barely rise to the chromosphere or higher
layers without interacting with the surrounding umbral fields \citep{2006A&A...453.1079J, 2018ApJ...854...92T}. Thus, in most
cases, the reconnection takes place in the lower atmosphere around the TMR, making observations of IJ-related TBs much more
frequent than those of TBs without IJs. Although relatively rare, the existence of the latter TB indicates that small-scale
emerging magnetic structures driven by vigorous convection upflows in the LB could sometimes reach a higher layer and then
reconnect with the surrounding fields, which is likely due to different magnetic environments of different LBs. In future work,
more cases of TBs without IJs should be investigated with spectral observations to examine whether this scenario agrees with
our results.

\subsection{TB Versus LW-I}\label{sect42}
As discussed above, an LW-I is characteristic of a continuous bright front oscillating with a stationary period, which is distinctly
different from a TB. Around the base of an LW-I, there is usually no obvious TB. It is natural and reasonable that TBs and LW-Is
are not closely related due to the difference between their driving mechanisms. Because IJs are always associated with TBs,
TBs, IJs, and LW-Is are further discussed together in the next subsection.

\subsection{IJ Versus LW-I}\label{sect43}
As shown in Figures \ref{fig2} and \ref{fig3}, IJ and LW-I are two vastly different activities. The following differences are obvious:
\begin{enumerate}
\item In the \emph{IRIS} 1400 {\AA} observations, LW-Is have a continuous bright front, and a bubble-like void forms in the wall
body without obvious brightenings detected around the base, while IJs are identified as emission features accompanied by a
footpoint TB.
\item An LW-I is rooted in the whole LB over a spatial range of about tens of Mm, while an IJ has a typical width of
approximately 1--2 Mm.
\item Compared to LW-Is with typical heights of 2--5 Mm, IJs have much larger heights of $\sim$10 Mm.
\item An LW-I rises with a projected velocity of $\sim$7 km s$^{-1}$, while an IJ has a projected velocity of $>$50 km/s.
\item The lifetime of an IJ is several minutes, while an LW-I could exist for a relatively long time, even up to several days.
\item Unlike the intermittence of an IJ, an LW-I has an almost stationary oscillating period of 4--5 minutes.
\end{enumerate}

These observational features support the interpretation of an LW-I as a wave-driven activity and that an IJ is caused by
intermittent reconnection. Studying two events with the coexistence of an LW-I and an IJ above the same LB,
\citet{2017ApJ...848L...9H} proposed that in some LBs where the leaked p-mode waves consistently driving the
oscillating LW-I above the whole LB, intermittent magnetic reconnection between the emerging fields and surrounding umbral
fields could occur in some sites and produce an upward IJ accompanied by the footpoint TB. Similar results were also reported
by \citet{2018ApJ...854...92T} through combining the observations from the GST and \emph{IRIS}, where strong evidence for two
components of the surge-like activity above LBs were found. Moreover, persistent short LW-Is are likely related to the upward
leakage of magneto-acoustic waves from the photosphere. In addition, \citet{2021ApJ...908..201L} found that some IJs launched
from a TB in an LB could break through the bright front of an LW-I, where small-scale bright blobs ejected from the oscillating
bright front are detected. These results reveal that although LW-Is and IJs (always associated with a TB) are caused by different
mechanisms and thus show two distinct groups of observational features, they could simultaneously appear in the same LB in some
events.

Recent high-resolution photospheric observations have revealed that in some LBs, filamentary and granular features could
be observed simultaneously at different parts \citep{2016A&A...596A...7S, 2019ApJ...882..175Y, 2020AA...642A..44H}. As a result,
the co-existence of different types of activities naturally raises a question: if such co-existence has anything to do with
different LB structures observed in the photosphere? It seems self-evident that the answer is yes because the photospheric
observations of LBs directly reflect the information of convective energy transport, which is one of the basic physical processes
driving the dynamic activities above LBs. We also point out that a comprehensive study about this topic in the future needs more
events which must satisfy the following two conditions: (1) the same LB should host more than one type of activities driven by
different mechanisms in the \emph{IRIS} or \emph{SDO} (E)UV imaging observations; (2) the LB should be simultaneously observed in
photospheric channels by some observatories with high spatial resolution, such as \emph{Hinode}, GREGOR, GST, NVST, and upcoming
DKIST. However, to our best knowledge, the number of such cases is very limited at the moment.

\subsection{TB Versus LW-II}\label{sect44}
\begin{figure*}
\centering
\includegraphics [width=0.82\textwidth]{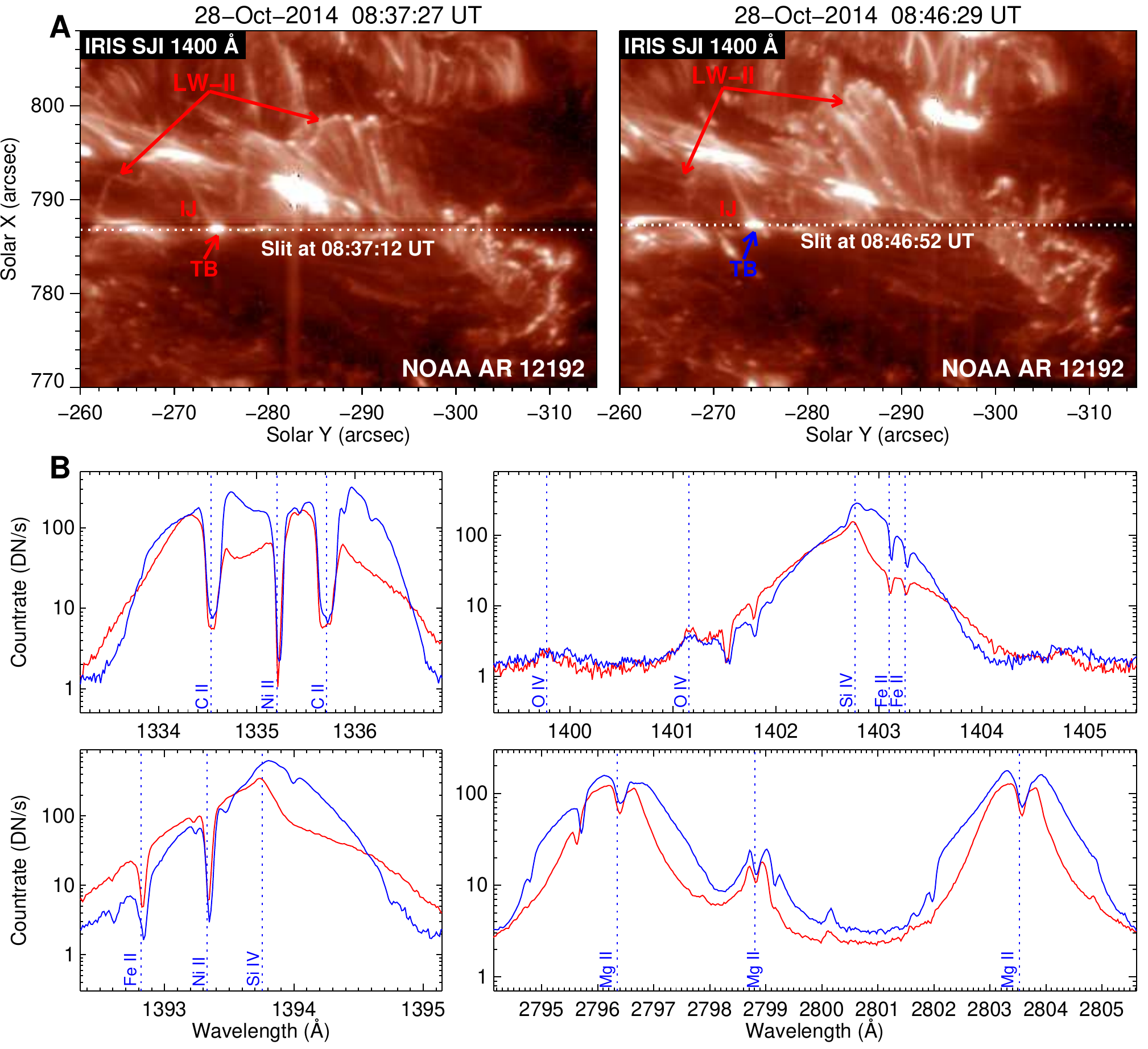}
\caption{\emph{IRIS} spectral observations of the base transient brightening of an LW-II above the sunspot LB in AR 12192 on 2014
October 28.
(A): \emph{IRIS} 1400 {\AA} images showing the LW-II and a notable brightening at the wall base at two time points.
(B): \emph{IRIS} line profiles of the TBs denoted by red and blue arrows in (A).
}
\label{fig7}
\end{figure*}

Because the LW-II shown in Figure \ref{fig4} was not crossed by the \emph{IRIS} slit, further spectral analysis is not possible. For
more information about the LW-II, in Figure \ref{fig7}, we show another typical LW-II above the sunspot LB of NOAA AR 12192 on 2014
October 28. This LW-II was crossed by the \emph{IRIS} slit several times and was investigated in detail by
\citet{2018ApJ...854...92T}. The successive \emph{IRIS} observations reveal that from 08:20 UT to 09:11 UT, this LW-II was rooted in
the whole LB and had an oscillating indented bright front with a height of $\sim$10 Mm in the 1400 {\AA} channel. TBs were also
frequently detected around the bases of this LW-II, above which ejections appeared as serried bright threads connecting to the bright
front. Panel (A) displays this LW-II around 08:06 UT and 08:46 UT, when the \emph{IRIS} slit crossed a recurrent TB. The profiles of
the C {\sc ii}, Si {\sc iv}, and Mg {\sc ii} lines of this TB measured at two time points are shown by red and blue curves in panel
(B). Both of them are remarkably enhanced and broadened at both wings of each emission line. Besides, signatures such as
superpositions of chromospheric absorption lines and change of the Mg {\sc ii} 2798.809 {\AA} line are also obvious here, just like
those of the IJ-related TBs shown in Figures \ref{fig2} and \ref{fig5}(B).

\begin{figure*}
\centering
\includegraphics [width=0.92\textwidth]{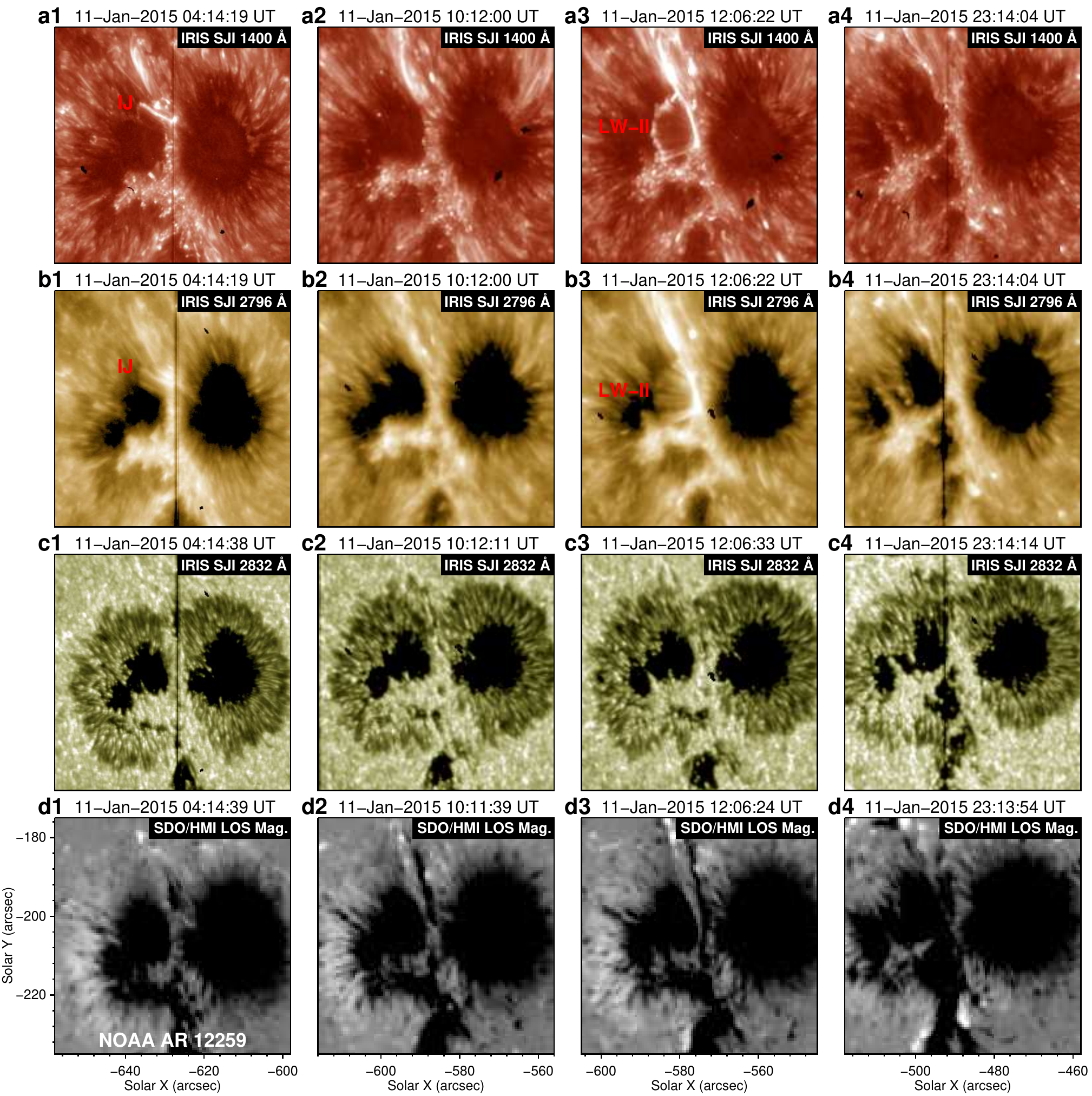}
\caption{Relation between the IJ and LW-II above the sunspot LB of AR 12259 on 2015 January 11. Sequences of \emph{IRIS} 1400 {\AA},
2796 {\AA}, 2832 {\AA} images, and \emph{SDO}/HMI LOS magnetograms in (a1)--(d4) display the evolution of the LB and dynamic
activities above.
}
\label{fig8}
\end{figure*}

Combining the imaging observational results shown in Section \ref{sect34} and the spectral results analyzed here, one can see that
different from the LW-I, the LW-II shows recurrent brightenings at the base along the whole LB, which are driven by frequent
reconnection occurring in the lower atmosphere. Then, plasma ejections are launched above these brightenings and appear as serried
bright threads, forming a wall-shaped structure in the \emph{IRIS} 2796 {\AA} and 1400 {\AA} images. It is obvious that the LW-II is
closely related to the brightenings as well as IJs. Further discussion of the relations between these three activities is presented
in the next subsection.

\subsection{IJ Versus LW-II}\label{sect45}
As shown in Figures \ref{fig2}, \ref{fig4}, \ref{fig5}(B), and \ref{fig7}, LW-IIs resemble IJs in many aspects:
\begin{enumerate}
\item Around the footpoint, TBs with similar spectral properties are detected.
\item The main body is seen in emission in the 2796 {\AA} and 1400 {\AA} channels but appears as a dark void region at the 171 {\AA}
and 94 {\AA} passbands.
\item Around the front, enhanced emissions can be dimly seen in the 1400 {\AA}, 171 {\AA} and 94 {\AA} images.
\item The ejections launched from TBs moved upwards very quickly and reached a height of $\sim$10 Mm.
\item The occurrence is intermittent without a nearly stationary period.
\end{enumerate}
It seems that the only difference between an IJ and an LW-II is their spatial ranges along the LB: an IJ has a typical width of about
1--2 Mm, while an LW-II has a spatial range over the whole LB.

To explore the relation between IJs and LW-IIs, in Figure \ref{fig8}, we display the evolution of the main sunspot of AR 12259 on
2015 January 11, where the LW-II exhibited in Figure \ref{fig4} and IJ shown in Figure \ref{fig2} are located in the same LB.
Around 04:00 UT, IJs were detected at the north end of this LB but were invisible in the next data set of \emph{IRIS} 1400 {\AA} and
2796 {\AA} around 10:00 UT. Around 12:00 UT, the LW-II appeared above the same LB but subsequently disappeared in the next
\emph{IRIS} observations around 23:00 UT. Due to the gap in the \emph{IRIS} observations, it is difficult to determine the
exact lifetime of these two activities. However, their disappearances in the next \emph{IRIS} observation period reveal that
their typical lifetimes could be a few hours at most. Simultaneous \emph{IRIS} 2832 {\AA} images and \emph{SDO}/HMI magnetograms
reveal that when the IJ and LW-II were observed, an elongated dark structure with negative magnetic polarity gradually intruded
into the LB from its north end, around which some TBs were detected in the 2832 {\AA} images.

These results indicate that the IJ and LW-II could essentially belong to the same category from the perspective of the driving
mechanism. The only difference between them is the spatial scope of reconnection occurring in the LB: the IJ is produced by
small-scale reconnection intermittently occurring at some localized sites of the LB, and the LW-II is caused by the frequent
reconnection occurring over the whole LB. Specifically, if the emerging magnetic field in the LB has a line-like shape or is
widely distributed in the entire bridge, the magnetic reconnection could frequently takes place in the lower atmosphere of the
LB over a large spatial range. Then, a dense row of IJs associated with footpoint TBs would be launched along the bridge during
some period of time. These IJs slam into the higher atmosphere and compress the local plasma in front, thus producing a wall-shaped
LW-II with a bright front. Because the reconnection cannot appear at all locations on the LB and at any time, the IJs should
be triggered randomly at a limited temporal and spatial scale. As a result, the bright front of an LW-II appears as a broken or
indented line rather than a continuous one like that of an LW-I.

Although the concept of LW-II is proposed in the present work for the first time, we speculate that such type of activity
actually has been reported in part of previous works. For example, as shown in Figure 2 of \citet{2016A&A...590A..57R}, fan-shaped
jets above a sunspot LB were clear in the SST and \emph{Hinode} observations, with lengths of about 7--38 Mm and an initial speed
of $\thicksim$100 km s$^{-1}$. A bright front was also observed ahead these fan-shaped jets and was interpreted by the compression
of coronal material due to the upward jet. Despite lack of \emph{IRIS} observations of these fan-shaped jets, we infer that such
piano-like jets could fall into the category of LW-IIs reported here according to their similar morphological, thermal,
and dynamical properties.
Similarly, \citet{2009ApJ...696L..66S} and \citet{2014A&A...567A..96L} reported a row of short jets above sunspot LBs and further
investigated the corresponding photospheric magnetic field by using \emph{Hinode} SOT and SP observations, respectively. We suggest
that the LW-II analyzed here by the \emph{IRIS} and \emph{SDO} observations could be the transition region and coronal counterparts
of the activity studied by \citet{2009ApJ...696L..66S} and \citet{2014A&A...567A..96L}.

\subsection{LW-I Versus LW-II}\label{sect46}
\begin{figure*}
\centering
\includegraphics [width=0.88\textwidth]{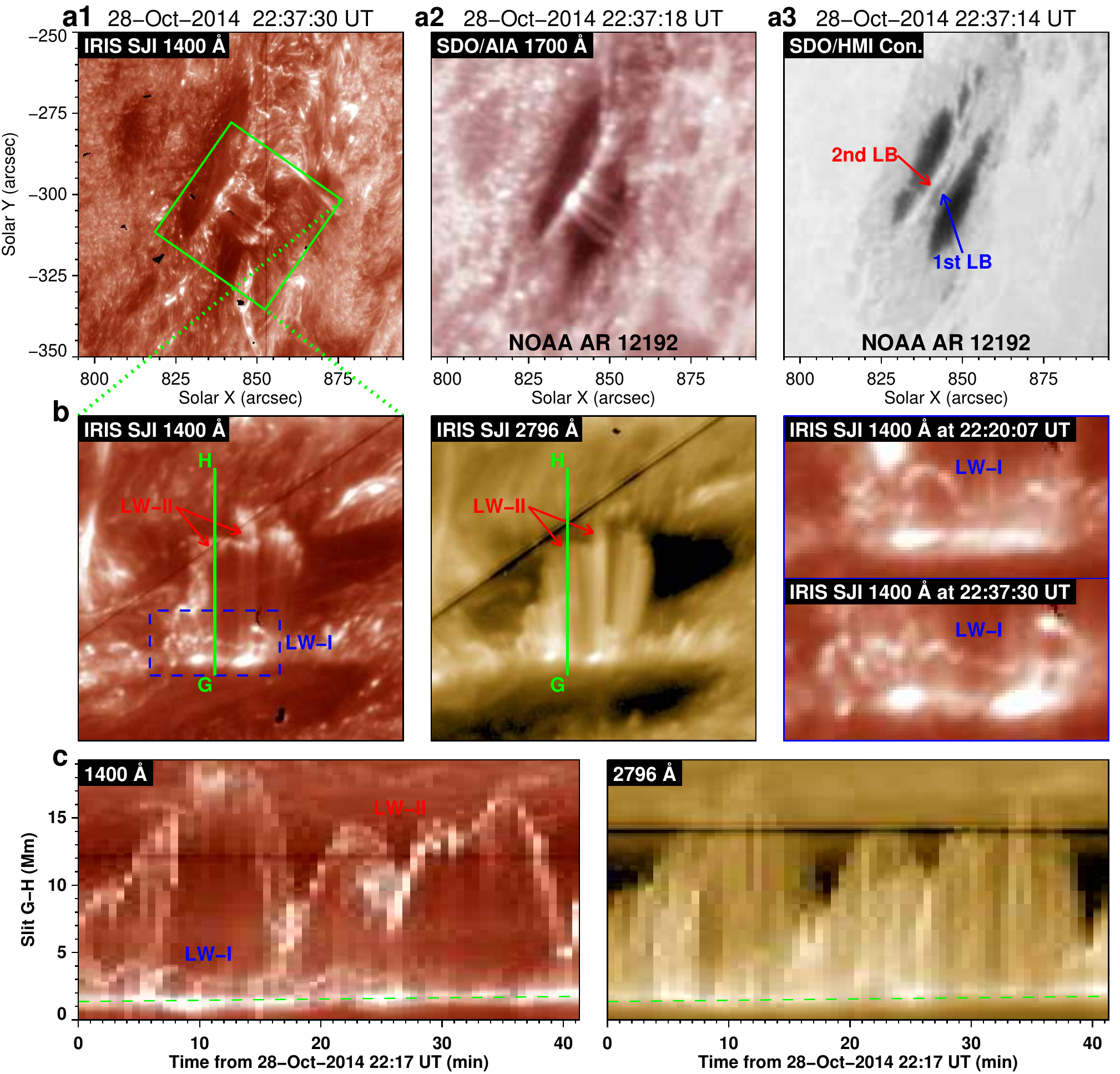}
\caption{Comparison between an LW-I and LW-II detected in the same sunspot of AR 12192 on 2014 October 28.
(a1)--(a3): Multi-wavelength images showing the LWs, LBs, and sunspot.
(b): Extended \emph{IRIS} 1400 {\AA} and 2796 {\AA} images clearly displaying the two types of LWs with distinct features.
(c): Time-distance plots along the ``G--H" slit in the \emph{IRIS} 1400 {\AA} and 2796 {\AA} channels.
An animation of the \emph{IRIS} 1400 {\AA} and 2796 {\AA} images covering the time from 22:17 UT to 22:58 UT on 2014 October 28
is available online. The animation's duration is 6 seconds.
}
\label{fig9}
\end{figure*}

LW-Is and LW-IIs as well as their relations have been analyzed in detail above. These two types of activities above LBs share a
notable feature of an oscillating bright front, and they have similar wall-like shapes, from which their names were derived.
However, they also show distinctly different properties and are driven by leakage of p-mode wave and magnetic reconnection,
respectively, which is why they are distinguished as type-I and type-II. Due to the different preferences for the forming
environment, these two types of LWs could not normally appear in the same sunspot LB. Figure \ref{fig9} displays a unique event
where an LW-I and LW-II were simultaneously detected in the same sunspot of AR 12192 on 2014 October 28. After checking the
evolution of this sunspot a few days before, we found that there were two LBs close and parallel to each other in the sunspot on
October 28. The west LB (1st LB in panel (a3)) formed at first and already produced an LW-I around October 25 (see Figure
\ref{fig4}). Over the following days, an additional LB (2nd LB) gradually formed besides, which was accompanied by dynamical
activities, such as TBs, IJs, and an LW-II. As this sunspot rotated to a location near the solar limb on October 28, it provides
an excellent viewpoint for us to observe the LW-I above the west LB and LW-II in the east LB in the same FOV. Figures \ref{fig9}(b)
and (c) show that the LW-I has a continuous oscillating bright front with a typical height of several Mm and a nearly stationary
period of several minutes. Meanwhile, the LW-II has a jagged bright front with a height of over 10 Mm, which oscillates without a
stable period and is accompanied by frequent ejections from the base.

Together with the results from the preceding sections, here we conclude the observational similarities and differences of the two
types of LWs as follows:
\begin{itemize}
 \item{Similarities}
   \begin{enumerate}
      \item Oscillating bright front in \emph{IRIS} 1400/1330 {\AA} observations.
      \item Spatial range over the entire LB.
    \end{enumerate}
  \item{Differences}
    \begin{enumerate}
      \item An LW-I has a continuous bright front, but the bright front of an LW-II is broken or indented.
      \item Around the base of an LW-I, there are usually no obvious TBs, but they recurrently appear in the whole bridge for an LW-II.
      \item Between the bright front and LB, a bubble-like void forms in an LW-I, while the main body of an LW-II consists of
      serried bright threads caused by the fast upward ejections.
      \item Compared to an LW-I with typical heights of 2--5 Mm, an LW-II has much a larger height of $\sim$10 Mm.
      \item An LW-I rises with a projected velocity of $\sim$7 km s$^{-1}$, while an LW-II has a projected velocity of $>$50 km s$^{-1}$.
      \item The lifetime of an LW-I ranges from several hours to several days, while an LW-II usually exists for several hours.
      \item An LW-I performs coherent oscillation with a nearly stationary oscillating period of 4--5 minutes, while the oscillation of
      an LW-II at two sites shows different patterns without constant period.
    \end{enumerate}
\end{itemize}

Based on these results, we speculate that in a developing sunspot LB, where large-scale magnetic flux emergence is driven by vigorous
convection upflows, magnetic reconnection will frequently occur along the whole bridge, and an LW-II could be a dominant activity. However,
an LW-I with a long lifetime and stable oscillating period should be located in an LB with a generally stable magnetic environment.

\subsection{Magnetic Field Comparison of the Four Types of Activities}\label{sect47}
\begin{figure*}
\centering
\includegraphics [width=0.98\textwidth]{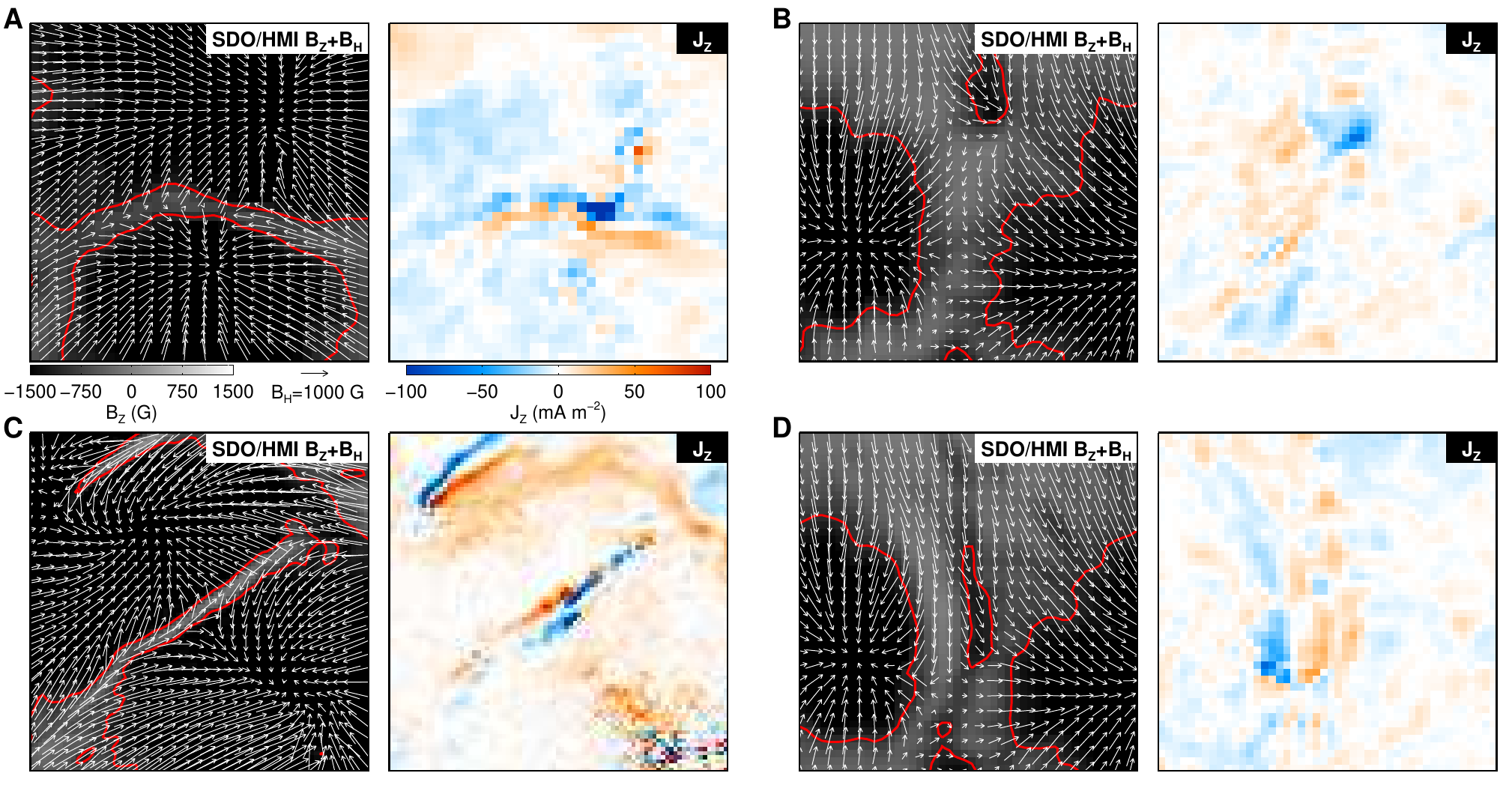}
\caption{Photospheric vector magnetic field and current density distribution of the LBs producing TB, IJ, LW-I, and LW-II.
A--D: Corresponding magnetic information of the different activities shown in Figures \ref{fig1}, \ref{fig2}, \ref{fig3}, and \ref{fig4},
respectively. Red curves mark -1000 G contour of the vertical component of vector magnetic field B$_{Z}$. White arrows represent the
horizontal component of vector magnetic field B$_{H}$.
}
\label{fig10}
\end{figure*}

Figure \ref{fig10} shows the photospheric vector magnetic field and the vertical current density distribution of the LBs
producing four types of activities, which are exhibited in Figures \ref{fig1}, \ref{fig2}, \ref{fig3}, and \ref{fig4}, respectively.
The vertical current density J$_{z}$ is calculated as follows:
\begin{equation}
J_{z} = \frac{(\nabla \times \textbf{B})_{z}}{\mu_{0}} = \frac{1}{\mu_{0}}(\frac{\partial B_{y}}{\partial x} - \frac{\partial B_{x}}{\partial y}),
\label{eq1}
\end{equation}
where $\mu_{0}$ is the magnetic permeability. Since J$_{z}$ is determined by the spatial gradient of B$_{H}$, it is natural that J$_{z}$
is large in LBs with relatively strong B$_{H}$, especially narrow ones, where B$_{H}$ significantly changes in a very small area.
In addition, the sustained emergence or intrusion of horizontal magnetic fields would also produce strong electric currents
\citep{2021A&A...652L...4L}. Figure \ref{fig10} reveals that LBs of these four events are all characterized by highly inclined
magnetic field and strong vertical current density J$_{z}$ around the sites where activities are detected, showing no obvious
differences with each other. However, Figures \ref{fig10}(B) and \ref{fig2} reveal that the IJ was located at the north end of
the LB, where a magnetic patch with strong and inclined fields was intruding into the LB. For the LW-II observed later in the same LB,
the LB at that time had been occupied by the intruding magnetic patch and showed a dynamical evolution. The LW-II was located between
the west part of umbra and the elongated intruding magnetic patch, where the vertical and horizontal components of the vector magnetic
field all changes sharply (see Figures \ref{fig10}(D) and \ref{fig4}). But for the TB and LW-I, there are no remarkable signals of
magnetic flux emergence or intrusion in the corresponding LBs (see Figures \ref{fig10}(A) and (C)), where the magnetic environment is
generally stable.

\subsection{Frequency Distribution Comparison of the Four Types of Activities}\label{sect48}
\begin{figure}
\centering
\includegraphics [width=0.48\textwidth]{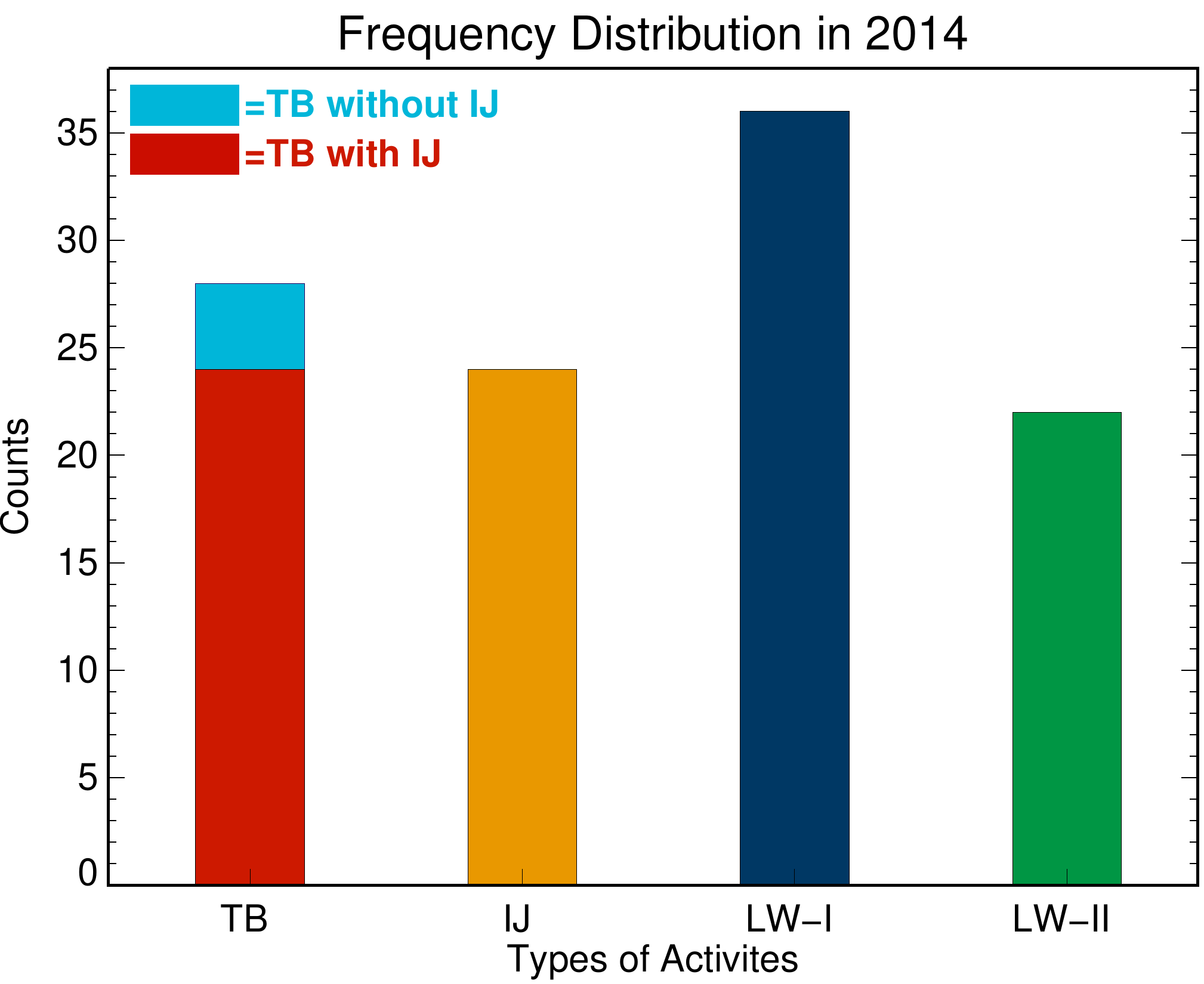}
\caption{Frequency distribution of the four types of activities above sunspot LBs in the whole year of 2014.
}
\label{fig11}
\end{figure}

To investigate frequency distributions of the different activities above sunspot LBs, we checked the \emph{IRIS} 1400/1330 {\AA}
observations of sunspot LBs in the whole year of 2014 and counted these activities. The counts of four types of activities follow
the two principles below: (1) An LW-I or LW-II above the same LB is counted only once, even if they lasts for a long time and are shown in
multiple \emph{IRIS} data sets. (2) IJs or TBs above the same LB are counted only once, even if they appear at different sites of this LB or
occur recurrently during the observing period. Figure \ref{fig11} reveals that in \emph{IRIS} 1400/1330 {\AA} observations of 2014, occurrence
frequencies of TB without IJ, IJ-related TB, IJ, LW-I, and LW-II are 4, 24, 24, 36, and 22, respectively. It is shown that $\thicksim86\%$ of
TBs are associated with IJs, and all IJs are accompanied by TBs. We also find that most of IJs are detected in LW-IIs, and sometimes observed
to be superimposed on some LW-Is or appear alone. LW-I is more frequently observed above sunspot LBs then LW-II.

\section{Summary}\label{sect6}
\begin{table*}
\caption{Various Activities above Sunspot Light Bridges in \emph{IRIS} Observations\label{t2}}
\centering
\begin{tabular}{c  c  c  c  c  c  c}   
\hline\hline
Activities  &  Size  &  Lifetime  &  Period  &  Velocity  &  Driving Mechanism  & References \\
\hline
 TB & Spot: 1--2 Mm      & several minutes & / & / & Magnetic reconnection &  \citet{2015ApJ...811..137T} \\
    & Elongated: 3--4 Mm &             &  &  &    &  \citet{2020AA...642A..44H} \\
\hline
 IJ & Height: $\gtrsim$ 10 Mm  & several minutes & / & $>$50 km/s & Magnetic reconnection &  \citet{2015ApJ...811..137T} \\
    &  Width: 1--2 Mm  &  -- several hours           &  &             &    &  \citet{2017ApJ...848L...9H,2020AA...642A..44H} \\
    &     &             &  &                &    &  \citet{2018ApJ...854...92T} \\
\hline
 LW-I & Height: 2--5 Mm  & several hours & 4--5 minutes & $\sim$10 km/s & Leakage of waves & \citet{2015ApJ...804L..27Y,2017ApJ...843L..15Y} \\
      &  Width: entire LB      &  -- several days  &       &        &             &  \citet{2015MNRAS.452L..16B} \\
      &     &             &  &                &    &  \citet{2016ApJ...829L..29H,2017ApJ...848L...9H} \\
      &     &             &  &                &    &  \citet{2017ApJ...838....2Z} \\
      &     &             &  &                &    &  \citet{2018ApJ...854...92T} \\
\hline
 LW-II & Height: $\sim$10 Mm  & several hours & / & $>$50 km/s & Magnetic reconnection &  The present work \\
       &  Width: entire LB           &           &  &                &           &    \\
\hline
\end{tabular}
\end{table*}

Based on the high-resolution observations from the \emph{IRIS}, we investigated various activities in sunspot LBs. Because these
activities display a wide variety of physical properties, we classified them into four distinct categories: TB, IJ, LW-I, and LW-II.
The observational characteristics, possible driving mechanisms, and recent works based on the \emph{IRIS} observations of the four
types of activities are summarized in Table \ref{t2}.

The four types of activities were further analyzed in comparison with each other. We summarize the results as follows:
\begin{enumerate}
\item In a sunspot LB, most observed TBs are associated with IJs and could be driven by magnetic reconnection between
emerging small-scale fields and surrounding umbral fields occurring in the lower atmosphere around the TMR. However, in
a few cases, the emerging magnetic structures might reach a higher layer and then reconnect with surrounding fields, producing
TBs without IJs due to the low local plasma density.

\item IJs and LW-Is above LBs are two vastly different activities and are caused by intermittent reconnection in lower
atmosphere and persistent upward leakage of magneto-acoustic waves from the photosphere, respectively. However, in some events,
these two mechanisms could simultaneously play roles in the LB and lead to the superimposition of IJs on an oscillating LW-I
at some sites.

\item The IJ and LW-II could essentially be the same category from the perspective of the driving mechanism. The only difference
between them is the spatial scope of reconnection driving them: an IJ is produced by small-scale reconnection intermittently
occurring at some localized sites of the LB, while an LW-II is caused by frequent reconnection occurring over almost the
entire LB.

\item When the emerging magnetic field has a line-like shape or is widely distributed along the entire LB, magnetic reconnection
could frequently take place, leading to recurrent TBs in the lower atmosphere of nearly the whole LB. Then, a dense row of IJs would
be launched above these TBs, which would slam into the higher atmosphere and compress the local plasma in front, thus producing the
bright front of a wall-shaped LW-II. Because the reconnection cannot simultaneously occur at all locations in the LB, the bright
front of the LW-II appears as a broken or indented line rather than a continuous one like that of an LW-I.

\item Although the LW-II is similar to the LW-I in the terms of oscillating bright front, they are essentially two different types
of activities from the perspective of the driving mechanism. The long lifetime and stable oscillating period of LW-I indicate that
the magnetic environment of an LW-I is generally stable. However, an LW-II is more likely to be detected in a dynamically
evolving sunspot LB, where large-scale magnetic flux emergence or intrusion are driven by vigorous magneto-convection and magnetic
reconnection frequently occurs along the whole bridge.
\end{enumerate}

Although sunspot LBs have been extensively studied for more than half a century, our understanding of this kind of special
sub-structure within a sunspot and its dynamics is still evolving, especially with the rapid development of solar
observatories in past decades. In the present work, we concentrated on the classification and comparison of the various activities
above LBs. However, the key factors that determine the occurrence of these activities and their possible relations to evolution of
a sunspot LB still remain unclear. To find these factors and eventually build a complete physical picture of the LBs and
associated activities, we need carefully investigate the physical linkage between the subphotospheric evolution and the atmospheric
dynamics above in the future works. More efforts ought to be focused on clarifying the regularities of these activities and their
exact relations to the general properties of sunspot LBs, especially from a statistical perspective. For instance, the analyzing
method applied in \citet{2018A&A...609A..73R} can comprehensively quantify the general physical properties of the LBs producing
different types of activities, and thus contribute to finding the key factors of interest.

In addition, there is another important topic that is not discussed in this paper: what happens above the LBs along the polarity
inversion line (PIL) of the delta-type sunspots or the LBs formed by coalescence of two pores with the same magnetic polarity? We have
noticed that in \emph{IRIS} 1400/1330 {\AA} images, there are no typical LB features (i.e., bright lane penetrating into the umbra) in
the delta-type sunspots. Instead, frequent flashing and dense bright loops dominate PIL region of the delta-type sunspot. The properties
of these activities above the LB of delta-type sunspots will be discussed in detail in an upcoming paper (Hou et al. 2022, in preparation).
As for the LBs through coalescence of two pores with the same magnetic polarity, the results of \citet{2019ApJ...882..110Z} could be
regarded as a reference: they found that the collision of two magnetic patches with the same polarity would launch fan-shaped activities
observed in H$\alpha$ channel. These fan-shaped activities ascended from the collision sites between the two patches and then decayed
within about 10 minutes. For achieving these topics, high-resolution data from other observatories, such as \emph{Hinode} and the
upcoming DKIST, will definitely be helpful. Besides, since we have learned a lot about LBs from the realistic sunspot simulations
\citep{2010ApJ...720..233C,2014ApJ...785...90R,2019ApJ...886L..21T}, with the advance of computational power, more insights into sunspot
LBs will be brought by further simulation works in the near future.

\acknowledgments
The authors are cordially grateful to two anonymous referees for their constructive comments and suggestions.
Y.J. Hou thanks Prof. Hui Tian and Dr. Qiangwei Cai for their helpful discussions. The data used here are courtesy of the \emph{IRIS}
and \emph{SDO} science teams. \emph{IRIS} is a NASA small explorer mission developed and operated by LMSAL with mission operations
executed at NASA Ames Research Center and major contributions to downlink communications funded by ESA and the Norwegian Space Centre.
\emph{SDO} is a mission of NASA's Living With a Star Program. The authors are supported by the Strategic Priority Research
Program of the Chinese Academy of Sciences (XDB41000000), the National Natural Science Foundation of China (11903050, 11790304,
11873059, 12073001, 12073042, and 11790300), the National Key R\&D Program of China (2019YFA0405000), the NAOC Nebula Talents Program,
Yunnan Academician Workstation of Wang Jingxiu (No. 202005AF150025), and Key Programs of the Chinese Academy of Sciences (QYZDJ-SSW-SLH050).
This work was also supported by JSPS KAKENHI Grant Nos. JP20KK0072, JP21H01124, and JP21H04492, and by the NINS Grant Nos. 01321802
and 01311904.



\bibliography{ref}{}
\bibliographystyle{aasjournal}

%
%

\end{document}